\documentclass[12pt]{article}

\usepackage{epsfig}
\usepackage{bbm}
\newcommand{\cA}{{\cal A}}

\newcommand{\cM}{{\cal M}}

\newcommand{\bi}{\bigskip}
\newcommand{\no}{\noindent}
\newcommand{\be}{\begin{eqnarray}}
\newcommand{\ee}{\end{eqnarray}}
\newcommand{\hk}{\hspace{0.1cm}}

\newcommand{\rk}{\right)}
\newcommand{\lk}{\left(}
\newcommand{\rke}{\right]}
\newcommand{\lke}{\left[}

\newcommand{\il}{\int\limits}

\begin{document}

\begin{center}
{\Large\bf Topology of Center Vortices\footnote{Supported by DFG
under grant number DFG-Re856/4-1.}}
\end{center}

\centerline{\rm H. REINHARDT}
\centerline{\itshape Institut f\"ur Theoretische Physik}
\centerline{\itshape Universit\"at T\"ubingen} 
\centerline{\itshape Auf der Morgenstelle 14}
\centerline{\itshape D-72076 T\"ubingen}

\begin{abstract}
The topology of center vortices is studied. For this purpose it is sufficient to
consider mathematically idealised vortices, defined in a gauge invariant way
as closed (infinitely thin) flux surfaces (in D=4 dimensions) which contribute
the $n^{th}$ power of a
non-trivial center element to Wilson loops when they are n-foldly linked to
the latter. In ordinary 3-space generic center vortices represent closed
magnetic flux loops which evolve in time. I show that
the topological charge of such a time-dependent vortex loop can be entirely
expressed by the temporal changes of its writhing number.
\end{abstract}

\no
\section{Introduction}
\label{dummy}
\bi

The vortex picture of confinement introduced already in the late 70s
\cite{[Hooft]} has only
recently received strong support from lattice calculations performed in the
so-called maximum center gauge \cite{[Del2]} where one fixes only the coset $G/Z$,
but leaves the center $Z$ of the gauge group $G$ unfixed. In this gauge the
identification of center vortices can be easily accomplished by means of the 
so-called center projection, which consists of replacing each link by its closest
center element. The so obtained vortex content is a physical property of the
gauge ensemble (in the sense of renormalization group invariance) \cite{[Lang]} 
and produces virtually the full string tension
\cite{[Ldel]}. Furthermore, the string tension disappears when the center
vortices are removed from the
Yang-Mills ensemble \cite{[Del2]}. This property of center dominance of the string tension
survives at finite temperature and the deconfinement phase transition can be
understood in a 3-dimensional slice at a fixed spatial coordinate as a
transition from a percolated vortex phase to a phase in which vortices cease to
percolate \cite{[Klang]}. Furthermore, by calculating the free energy of center vortices it has
been shown that the center vortices condense in the confinement
phase \cite{[Kova]}. It has also
been found on the lattice that if the center vortices are removed from the
gauge ensemble, chiral symmetry breaking disappears and all field configurations
belong to the topologically trivial sector \cite{[Forc]}. 
\medskip

In $D=4$ center vortices are closed magnetic flux sheets. In ref. \cite{[Eng]} it
was shown that the topological charge (Pontryagin index) of center vortex
sheets is given by their intersection number\footnote{Note that in 4 dimensions 
2-dimensional sheets
generically intersect in points.} 
, see also refs. \cite{[Reinh2]} ,
\cite{[Corn]}. Based on
this result the topological susceptibility was calculated for a random vortex
model \cite{[Eng3]} in ref. \cite{[Eng2]} and for the center projected vortex
ensemble in ref. \cite{[Ber2]}.
\medskip

In the present paper I
study the topology of generic center vortices, which represent (in general time-
dependent) closed magnetic flux loops, and express their topological charge in
terms of the topological properties of these loops. I will show that
the topological charge of generic center vortices is given by the temporal change of
the writhing number of the magnetic flux loops.
\medskip

The organisation of the paper is as follows: In sect. 2 I give a topological 
definition of center vortices in continuum
Yang-Mills theory by means of their effect on Wilson loops and give explicit
representations of the gauge potential of center vortices. In particular I
construct the gauge potential of generic center vortices representing, at a
fixed time, closed
magnetic flux loops. In sect. 3 I discuss the various types of singular
points of vortex sheets carrying non-zero topological charge, like intersection
points or twisting points. In sect. 4 I
express the topological charge of generic center vortex loops by their
writhing number. I illustrate the obtained result by a means of an example in
sect. 5. Finally a short summary and some concluding remarks are given in
sect. 6. 

\no
\section{Center vortices in continuum Yang-Mills theory}
\subsection{Definition of center vortices}
\bi

\no
In D-dimensional continuum Yang-Mills theory center vortices are localised gauge 
field configurations which
carry flux, which is concentrated on  $D=2$ dimensional closed hypersurfaces (i.e.
on closed sheets in D=4 dimensions or on closed
loops in D=3 dimensions).
Furthermore, their flux is quantized such that they contribute a non-trivial
center element to a (large) Wilson loop in the fundamental representation, when 
they are non-trivially linked to the latter.
To be more precise, if we assume that the flux is concentrated on the closed
hypersurface $\partial \Sigma$, then we can define the gauge potential $A_\mu$ of a center vortex
in D-dimensions by means of its effect on a Wilson loop by the relation

\be
\label{G1}P e^{- \oint\limits_C dx_\mu A_\mu (\partial \Sigma)} = 
Z^{L (C, \partial \Sigma)} \hk ,
\ee

where $A_\mu = A_\mu ^a T_a$ with $T_a$ being the 
generators\footnote{We use antihermitean generators of the gauge group.}
of the gauge group in the fundamental representation and $P$ denotes path 
ordering. Furthermore $Z$ denotes a non-trivial center element of the gauge group and 

\be
\label{G2}L (C, \partial \Sigma) = \oint\limits_C d x_\mu 
\oint\limits_{\partial \Sigma} d^{D - 2}
\tilde{\sigma}_{\mu \nu} \partial^x_\nu D^{(D)} 
\lk x - \bar{x} (\sigma) \rk
\ee

is the linking number between the Wilson loop C and the closed vortex
hypersurface $\partial \Sigma$.
Here $\bar{x}_\nu (\sigma)$ denotes a parametrisation of the vortex sheet
$\partial \Sigma$ with parameters $ \sigma = \lk \sigma_1 ,..
\sigma_{D-2} \rk $ and 
$ d^{D-2} \tilde{\sigma}_{\mu \nu} = 
\frac{1}{(D-2)!} \epsilon_{\mu \nu_{\alpha_1}} ..._{\alpha_{D-2}} d^{D-2} 
\sigma_{\alpha_1 ...\alpha_{D-2}}$  is 
the dual of the surface element 

\be
\label{G3}d^{D-2} \sigma_{\alpha_1 ...\alpha_{D-2}} = \varepsilon_{a_1} ..._{a_{D-2}}
\partial_{a_1} x_{\alpha_1} (\sigma)... \partial_{a_{D-2}} x_{\alpha_{D-2}}
(\sigma),  \hspace{0.5cm} \partial_{a_i} = \frac{\partial}{\partial
\sigma}_{a_i} \hk .
\ee

Furthermore, $D^{(D)} \lk x - \bar{x} (\sigma) \rk$ denotes the Green function 
of the $D$ dimensional Laplacian 

\be
\label{G4}- \partial^2 D^{(D)} (x) = \delta^{(D)} (x) \hk .
\ee

For the gauge group SU(N) the center Z(N) is given by the N different
$N^{th}$ roots
of unity

\be
\label{G5}Z (k) = e^{i \frac{2\pi}{N} k} \mathbbm{1} ,\hspace{1cm} k = 0,1 ... N-1 
\ee

where $\mathbbm{1}$ denotes the N-dimensional unit matrix. Like all group
elements, the center elements $Z(k)$ can be represented as exponentials of Lie algebra-valued
vectors 

\be
\label{G6}
Z(k) = e^{-E(k)} ,  \qquad E(k) = E^a(k) T_a .  
\ee

Up to a factor of $2\pi$ these vectors $E(k)$ are given by the co-weights $\mu
(k) \lk E(k) = 2\pi \mu (k) \rk $, which live entirely in the Cartan subalgebra
and define the corners of the fundamental domain of the $SU(N)$ algebra. They
are dual to the simple roots and satisfy the relation 

\be
\label{G8}
e^{-\hat{E} (k)} = \mathbbm{1} ,\hspace{1cm} \hat{E} (k) = E^a(k) \hat{T}_a
\ee

where $\hat{T}_a$ denotes the generators of the gauge group in the adjoint
representation, $\left( \hat{T}_a \right) {}^{bc} = f^{bac}$.
\medskip 

Given the property $Z(k)^N = \mathbbm{1}$, eq.~(\ref{G1}) implies that the Wilson 
loop is a vortex
counter mod. N. Furthermore, comparing eqs.~(\ref{G6}) and (\ref{G8}) it is seen
while the quarks,
living in the fundamental representation, do see center vortices the gluons,
living in the adjoint representation, are center blind and do not see center
vortices. To gluons center vortices look like (unobservable) Dirac strings. (A
center vortex has no effect on a Wilson loop in the adjoint representation).
\medskip

 From the definition of the center vortices (\ref{G1}) it is clear that the different
(non-trivial) center elements $Z(k)$ (\ref{G5}) are associated with different types
of center vortex flux. For the gauge group $SU(2)$ there is only one single
non-trivial center element $Z(1)=-1$ and hence only one type of center vortices.
For $SU(N>2)$ there are $N-1$ non-trivial center elements giving rise to $N-1$
different center vortex fluxes. Since

\be
\label{zzz} Z(k) = Z(1)^k
\ee

fusion and fission of center vortices are possible. However, due to the relation
(\ref{zzz}) a center vortex with flux corresponding to the center element $Z(k)$ 
can be regarded as a bunch of
$k$ center vortices, each associated with the basic non-trivial center
element $Z(1)$. For topological consideration it is therefore sufficient 
to consider only center vortices
associated with the basic element Z(1). For this reason it is also sufficient
to consider the gauge group $SU(2)$, to which we will confine myself later on.

\no
\subsection{The gauge potential of center vortices}
\bi

In principle, the gauge potential of a center vortex will have non-abelian
components. However, since the center of a gauge group is entirely in the Cartan
subgroup, one can choose the  gauge potential of center vortices in the Cartan
subgroup. 
\medskip

 From the definition of center vortices (\ref{G1}) and the
definition of the linking number (\ref{G2}), one can read off that a specific
realisation of
the gauge potential of a center vortex living on the closed hypersurface $\partial \Sigma$ is given by
$(E = E(k))$
\be
\label{G9} a_\mu (\partial \Sigma, x) = - E \il_{\partial \Sigma} d^{D - 2}
\tilde{\sigma}_{\mu \nu} \partial^x_\nu D \lk x - \bar{x} (\sigma) \rk  
\ee

where I have assumed the minimal flux necessary to produce in eq.~(\ref{G1})
the desired center element. Indeed, this gauge potential represents a center vortex, whose flux

\be
\label{N1} {\cal F}_{\mu \nu} \lk \partial \Sigma ,x \rk = \partial_\mu a_\nu \lk \partial \Sigma
,x \rk - \partial_\nu a_\mu \lk \partial \Sigma ,x \rk 
\ee

is entirely located on the infinitesimally thin vortex hypersurface $\partial
\Sigma$ 

\be
\label{F2}{\cal F} _{\mu \nu} \lk \partial \Sigma ,x \rk = E \int\limits_{\partial \Sigma} d^{D-2}
\tilde{\sigma}_{\mu \nu} \delta^{(D)} \lk x - \bar{x} (\sigma) \rk.
\ee

In this respect, this gauge
potential (\ref{G9}) represents a  mathematical idealisation of a center vortex. 
From the physical point of
view, center vortices are extended objects, where the flux is smeared out in the
transversal direction perpendicular to the vortex hypersurface $\partial \Sigma$. In other words,
physical center vortices have a finite thickness. This thickness can be measured on the
lattice and is obtained to be typically of the order of one fm \cite{[Del2]}. For
the present considerations, where I
concentrate on the topological properties of center vortices, the finite
transversal
extension of these vortices is completely irrelevant. In fact, one can assume an
arbitrary shape function for the transversal extension of the center vortices, which, 
however, drops out from
the topological charge (Pontryagin index), which I am interested in here, see
ref. \cite{[Eng]}. So in the following I will use the 
mathematical idealisation of a center vortex gauge
potential defined by eq.~(\ref{G9})\footnote{In {\bf dynamical} considerations
it is unavoidable to use physical vortices of finite ``thickness''. This is because
the mathematically idealised thin vortices have infinite Yang-Mills
action, see also ref. \cite{[Eng]}. Furthermore, thick vortices also contribute non-trivially
to adjoint Wilson loops and can account for the so-called Casimir scaling
\cite{[Fab]}.}. 
As a side remark, I mention that this
gauge potential fulfils the Lorentz gauge $\partial_\mu a_\mu (\partial \Sigma,
x) = 0$ (except on $\partial \Sigma$). 
\bi

\no
Due to the fact that the linking number $L (C, \partial \Sigma)$ between a loop
$C$ and a closed surface $\partial \Sigma$ equals the intersection number

\be
\label{G10}
I (C, \Sigma) = \oint_C d x_\mu \il_\Sigma d^{(D-1)} \tilde{\sigma}_\mu \delta^{(D)} \lk x -
\bar{x} (\sigma) \rk = L (C, \partial \Sigma) 
\ee

between the loop $C$ and the $D-1$-dimensional hypersurface (volume) $\Sigma$ enclosed by the sheet
$\partial \Sigma$, one can give an alternative representation for the gauge
potential of an idealised center vortex in the form

\be
\label{G11}
\cA_\mu (\Sigma, x) = E \il_\Sigma d^{D - 1} \tilde{\sigma}_\mu \delta^{(D)} \lk x -
\bar{x} (\sigma) \rk \hk .
\ee

It is worth mentioning that this form of the gauge potential is precisely the
continuum analogue of the gauge configurations arising on the lattice after
center projection \cite{[Eng]}. This type of vortex fields have been referred to in
\cite{[Eng]} as $ideal$ center vortex, while the gauge potential defined
by eq.~(\ref{G9}) has been referred to as $thin$ center vortex. While the thin vortex potential $a \lk
\partial \Sigma ,x \rk $ (\ref{G9}) manifestly depends only on the boundary $ \partial
\Sigma $ where the flux associated with the vortex is located, the ideal vortex
gauge potential $ \cA (\Sigma ,x)$ (\ref{G11}) is defined
on the $(D-1)$-dimensional hypersurface $\Sigma$. For fixed boundary $\partial
\Sigma$ (i.e. for fixed center vortex flux) different choices of $\Sigma$
correspond to different choices of the $Z(N)$ gauge. Indeed, consider two
(D-1)-dimesional hypersurfaces, $\Sigma_1$ and $\Sigma_2$, having the same boundary $\partial
\Sigma _1 = \partial \Sigma _2$. Then\footnote{Here $\lk -\Sigma \rk $ denotes
the hypersurface resulting from $\Sigma$ by reversing its orientation.}

\be
\label{11}\Sigma _1 \cup \lk -\Sigma _2 \rk = \partial \cM
\ee

represents the boundary of a D-dimensional hypersurface (volume) $\cM$. The
center gauge transformation $g (\cM , x) \in Z (N)$ which converts $\cA \lk
\Sigma_1 ,x \rk $ into $\cA \lk \Sigma _2 ,x \rk $ is given by

\be
\label{458}g \lk \cM ,x \rk = \exp \lk -E\chi \lk \cM ,x \rk \rk
\ee

where

\be
\label{22}
\chi \lk \cM ,x \rk = \int\limits_{\cM} d ^D \tilde{\sigma} \delta
^{(D)} \lk x -
\bar x (\sigma)\rk = 
\left\{ \begin{array}{ccc}
1 & , & x \in \cM\\
0 & , & \mbox{otherwise}
\end{array} \right.
\ee

is the characteristic function  of $\cM$. In fact, with (\ref{458}) one has 

\be
\cA _\mu \lk \Sigma _1 ,x \rk ^{g(\cM , x )} & \equiv &  g \lk \cM , x \rk \cA _\mu
\lk \Sigma _1 , x \rk g^\dagger \lk \cM ,x \rk + g \lk \cM ,x \rk \partial _\mu
g^\dagger \lk
\cM ,x \rk \nonumber\\ 
& = & \cA_\mu \lk \Sigma_1 ,x \rk + E \partial _\mu \chi \lk \cM ,x \rk =
\cA_\mu \lk \Sigma_2 ,x \rk , 
\ee

where I have used 

\be
\partial_\mu \chi \lk \cM ,x \rk = - \int\limits_{\partial \cM} d^{D-1}
\tilde{\sigma}_\mu \delta^{(D)} \lk x-\bar x (\sigma)\rk
\ee

which follows from (\ref{22}) by means of Gauss' theorem.
\medskip

Since $a_\mu \lk \partial \Sigma ,x \rk $ (\ref{G9}) and $ \cA_\mu \lk \Sigma ,x
\rk $ (\ref{G11})
both produce the same Wilson loop, they have to be gauge equivalent. In fact one can
show \cite{[Eng]} that

\be
\label{G12}
\cA_\mu \lk \Sigma ,x \rk = a_\mu \lk \partial \Sigma, x \rk + V \lk \Sigma ,x \rk
\partial_\mu V \lk \Sigma ,x \rk 
\ee

where the gauge transformation 

\be
\label{G13}
V \lk \Sigma ,x \rk = \exp \lk -E \Omega \lk \Sigma ,x \rk \rk 
\ee

is defined by the solid angle $\Omega \lk \Sigma ,x \rk $ subtended by the
($D-1)$-dimensional hypersurface $\Sigma $ from the point $x$ 

\be
\label{G14}
\Omega \lk \Sigma ,x \rk = \int \limits_\Sigma d^{D-1} \tilde{\sigma}_\mu
\partial^x _\mu D \lk x - \bar{x}(\sigma) \rk . 
\ee

Furthermore, one can show that the thin
vortex gauge potential (\ref{G9}) is just the transversal part of the ideal center vortex
potential (\ref{G11})

\be
\label{G15}
a_\mu (\partial \Sigma, x) = P_{\mu \nu} \cA_\nu (\Sigma,
x), \hspace{1.0cm} P_{\mu \nu} = \delta_{\mu \nu} - \frac{\partial_\mu
\partial_\nu}{\partial^2}.
\ee

In the following
I will refer to both $a_\mu \lk \partial \Sigma \rk $ and $ \cA_\mu \lk \Sigma \rk
$ as gauge potential of an (ideal and thin) center vortex since both potentials
describe the same mathematically idealized, infinitesimally thin flux sheet. To
distinguish between these two potentials I will refer to $ a_\mu \lk \partial \Sigma ,x \rk $ and $\cA_\mu \lk \Sigma ,x
\rk $ as regular and singular gauge, respectively. Note that $a_\mu \lk \partial
\Sigma , x \rk $ is a well behaved function of $x$
except on the vortex sheet $\partial \Sigma$ itself, while $\cA _\mu \lk \Sigma
,x \rk$ has support only on the hypersurface $\Sigma$ where it diverges. 

\no
\subsection{The flux of center vortices}
\bi

Whether the flux of a center vortex (\ref{F2}) is electric or magnetic, or both depends on
the position of the ($D - 2$)-dimensional vortex surface $\partial \Sigma$ in $D$
-dimensional space. For definiteness let us consider a D-dimensional Euclidean
space-time manifold. It is spanned by $(D-1)$ spatial and one temporal basis
vector. A vector is called $spatial$ if it is spanned entirely from the spatial basis
vectors. Similarly, a vector parallel to the temporal basis vector is referred to
as $temporal$. In this space
$(d\le D)$-dimensional hypersurfaces are spanned by $d$ tangent vectors. A
($d\le D$)-dimensional infinitesimal hypersurface element
$\delta \Sigma^{(d)}$ is spanned by $d$ linear independent tangent vectors. If
all $d(\le (D-1))$ tangent vectors are spatial the hypersurface element
is called $spatial$, $\delta \Sigma^{(d)}_s$, if one tangent vector is temporal it is
called $temporal$, $\delta \Sigma^{(d)}_t$. A hypersurface $\Sigma$ completely
built up from $spatial$ or $temporal$, respectively, hypersurface elements is
referred to as $spatial$ $\Sigma_s$ or $temporal$ $\Sigma_t$, respectively. Note that
generic vectors and hypersurfaces are neither temporal nor spatial. For definiteness 
let us confine ourselves in the following to D=4 space-time
dimensions.

\subsubsection{Spatial vortex surfaces}

\begin{figure}
\centerline{
\epsfysize=14.0cm
\epsffile{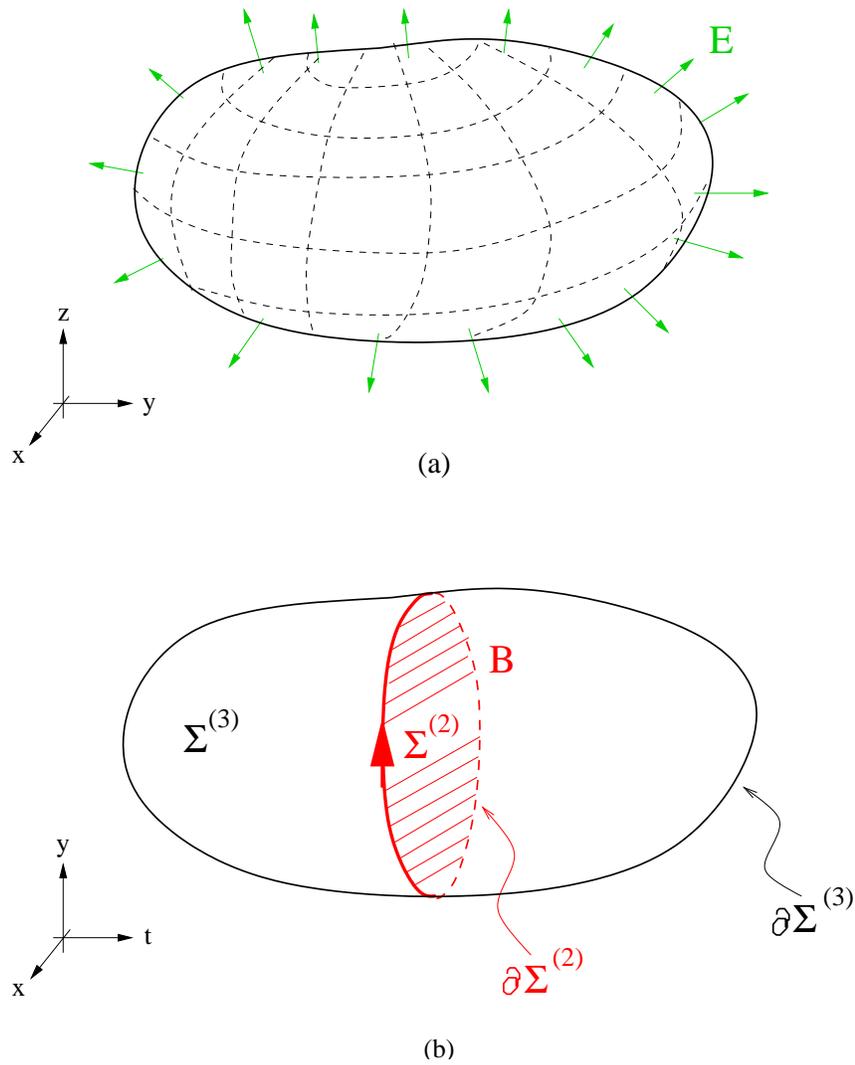}
}
\caption{Illustration of (a) a (non-generic) electric center vortex and (b) a
magnetic center vortex in ordinary 3-space. While the electric vortex
sheet (a) exists only at a single time-instant the magnetic vortex (b) evolves
in time.}
\label{fig1}
\end{figure}

Consider a center vortex defined by  a
purely spatial boundary $\partial \Sigma_s$ (which is the
boundary $\partial \Sigma_s$ of a purely spatial 3-dimensional volume
$\Sigma_s$). It carries only electric
flux, which is directed normal to the vortex surface $\partial \Sigma_s$, see
Fig. \ref{fig1} (a). Indeed, for a purely spatial 3-volume $\Sigma_s$ the spatial 
component of the gauge
potential vanishes

\be
\label{G16}
\vec{\cA} \lk \Sigma_s , x \rk = 0
\ee

as follows from eq.~(\ref{G11}) and the definition of the dual surface element $d^3
\tilde{\sigma}_\mu = \frac{1}{3} \epsilon_{\mu \alpha \beta \gamma} d^3
\sigma_{\alpha \beta \gamma}$. Consequently the $\vec{B}$-field of such vortices
vanishes.  For such vortices the temporal part of the gauge potential becomes
(with $d^3 \tilde{\sigma}_0 = d^3 x$)

\be
\label{G17}
\cA_0 \lk \Sigma_s, x \rk = E \delta \lk x_0 - \bar{x}_0
\rk \chi \lk \Sigma_s , \vec x\rk ,
\ee

where $\bar{x}_0$ is the time instant at which the purely spatial center vortex surface $\partial
\Sigma_s$ exists, and $ \chi \lk \Sigma ,\vec{x} \rk $ is the characteristic function 
of the (spatial) 3-volume $\Sigma$ defined by eq.~(\ref{22}) with $D=3$. The gauge
potential (\ref{G17}) gives rise to an electric field

\be
\label{G19}
\vec{E} \lk \partial \Sigma, x \rk = - \vec{\partial} \cA_0 (\Sigma, x) = - \delta
\lk x - \bar{x}_0 \rk E \, \vec{\partial} \, \chi_\Sigma \lk \vec{x} \rk \hk ,
\ee

which is concentrated on the spatial surface $\partial \Sigma$ and is normal to
it. The purely spatial vortex surfaces $\partial \Sigma_s$ existing at a single
time-instant can be considered as pathological.
\medskip

The generic case will be closed vortex sheets $\partial \Sigma$ evolving in time
and at a fixed time these vortex sheets represent closed loops $C$ of magnetic
flux, which is tangential to the vortex loop (see fig. \ref{fig1} (b)). 

\subsubsection{Temporal vortex surfaces}

For temporal hypersurfaces $\Sigma_t$
the temporal part of the gauge potential (\ref{G11}) has obviously to
vanish

\be
\label{G20}
\cA_0 (\Sigma_t , x)= 0
\ee

as follows again from the property of the dual volume element $d
\tilde{\sigma}_\mu$.
\medskip

Consider now the spatial part $\vec{\cA} \lk \Sigma _t ,x \rk $ (\ref{G11}) of
a center vortex in the singular gauge defined on a 3-dimensional temporal
hypersurface $\Sigma_t ^{(3)}$ in D=4. $\Sigma_t ^{(3)}$ is swept out by the time-evolution of
a 2-dimensional open surface (disc) $\Sigma^{(2)}(t)$ in ordinary
3-space $\mathbbm{R}^3$ (see fig. \ref{fig1} (b)), i.e. at a fixed time
$t, \Sigma_t ^{(3)}$ represents
a (open spatial) disc $\Sigma_s ^{(2)} (t)$.
In fact, by definition, a purely temporal hypersurface contains a temporal
tangent vector. Therefore one can identify one of the parameters $\sigma_{i = 1, 2, 3}$
characterising the hypersurface $\Sigma_t ^{(3)}$ in 4-dimensional space with time,
e.g. $\sigma_1 = \bar{x}_0 = \bar{t}$. Using

\be
\label{G21}\il_{\Sigma_t ^{(3)}} d^3 \tilde{\sigma}_k = 
\frac{1}{3!} \epsilon_{k \alpha \beta \gamma} \il_{\Sigma_t ^{(3)}} d^3
\sigma_{\alpha \beta \gamma} = \frac{1}{3!} \int d \bar{t}
\il_{\Sigma^{(2)} (\bar{t})} 3 \epsilon_{k 0 i j} d^2 \sigma_{i j} = \int d
\bar{t} \il_{\Sigma^{(2)} (\bar {t})} d^2 \tilde{\sigma}_k, 
\ee

where $d^2 \tilde{\sigma}_k = \frac{1}{2} \epsilon_{kij} d^2 \sigma_{i j}$, 
and $\delta^{(4)} \lk x - \bar{x} (\sigma) \rk = \delta \lk t - \bar{t} \rk
\delta^{(3)} \lk x - \bar{x} \lk t , \sigma_1, \sigma_2 \rk \rk$ the gauge
potential of a temporal center vortex (in singular gauge) becomes

\be
\label{G22}
\cA_k \lk \Sigma_t ^{(3)}, \vec{x} ,t \rk = E 
\il_{\Sigma_s ^{(2)} (t)} d^2 \tilde{\sigma}_k
\delta^{(3)} \lk  \vec{x} - \bar{x} \lk t, \sigma_2, \sigma_3 \rk \rk
\hk .
\ee

This gauge potential (\ref{G22}) gives rise to a magnetic field $\vec B = \vec
\partial \times \vec \cA$, which by means of Stoke's theorem is obtained as 

\be
\label{ss}B_i \lk \partial \Sigma_s ^{(2)} (t) , \vec x \rk = E
\oint\limits_{\partial \Sigma_s
^{(2)} (t)} dx'_i \delta^{(3)}\lk\vec x -\vec{x'} (\sigma) \rk .
\ee

Indeed this field is directed tangential to the closed loop $\partial \Sigma
^{(2)}$. Furthermore, due to its time-dependence the vector gauge potential
(\ref{G22}) generates also an electric field

\be
\label{sss}\vec E = \partial _t \vec{\cA} \lk \Sigma_s ^{(2)} ,t \rk
\ee

which is normal to the surfaces $\Sigma^{(2)} (t)$.
\medskip

A comment is here in order concerning the time-depence of the open spatial disc
$\Sigma_s ^{(D-2)} (t)$ which during its time-evolution traces out the temporal
hypersurface $\Sigma_t ^{(D-1)}$. During its time-evolution $\Sigma_s ^{(D-2)}
(t)$ changes only near its boundary $\partial \Sigma _s ^{(D-2)} (t)$ (which
represents the center vortex at a fixed time t) as is illustrated for the
$D=2+1$ dimensional case in fig. \ref{fig2}: The temporal hypersurface $\Sigma^{(2)}_t$ given by the
mantle of the cylinder is swept out by the time-evolution of an open (spatial)
string $\Sigma_s ^{(1)} (t)$ whose boundary $\partial \Sigma _s ^{(1)}$ (given by
its two endpoints) represents the vortex at a fixed time $t$. At the initial
time $t = \bar t _i$ when the vortex(-anti-vortex-pair) 
is created $\Sigma _s
^{(1)} \lk \bar t _i \rk$ is given by a single point (the position of vortex and
antivortex). As the time increases $\Sigma _s ^{(1)}$ becomes an open string
which growths at its end points 
until $t = \bar t _f$ where it becomes a closed loop, which has no boundary and
therefore carries no vortex. (The vortex disappears at $t = \bar t _f$). The closed spatial loop $\Sigma _s ^{(1)} \lk t
\ge \bar t _f \rk $ remains unchanged up to the time $t = \bar t _s$ where it
merges to the spatial surface $\Sigma _s ^{(2)}$. The sudden disappearence of
$\Sigma _s ^{(1)} (t)$ at $t = \bar t _s$ induces an electric field

\be
\vec E = \partial _t \vec{\cA} \lk \Sigma _s ^{(1)} (t) \rk 
\ee

which compensates the electric field

\be
\vec E = -\vec{\partial} \cA _0 \lk \Sigma _s ^{(2)} \rk
\ee

which is carried by the spatial (hyper-)surface $\Sigma_s ^{(2)}$ at its
boundary $\partial \Sigma _s ^{(2)} = \Sigma _s ^{(1)} \lk \bar t _s \rk$.

\subsubsection{Generic vortex surfaces}

\begin{figure}
\centerline{
\epsfysize=20.0cm
\epsffile{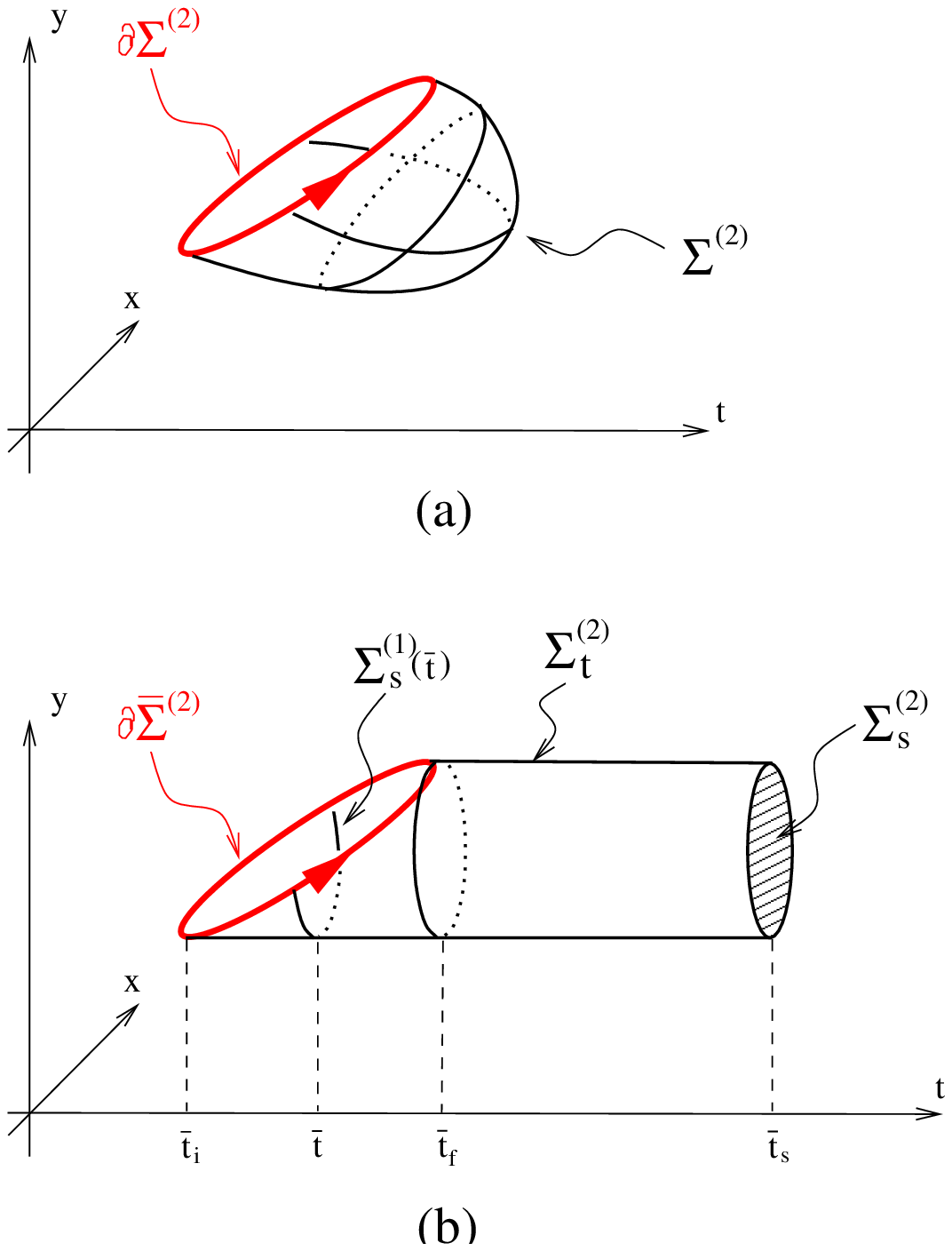}
}
\caption{(a)Illustration of a mathematically idealised center vortex loop
$\partial \Sigma^{(2)}$ as boundary of an open 2-dimensional surface
$\Sigma^{(2)}$, which is neither temporal nor spatial. (b) The same vortex loop
$\partial \Sigma ^{(2)}$ but with a different hypersurface $\bar \Sigma ^{(2)} =
\Sigma _t ^{(2)} + \Sigma _s ^{(2)}$, which
has been chosen as the surface of an open cylinder whose mantle $\Sigma_t$ is
temporal while its face $\Sigma_s$ is spatial. Also shown are fixed time $\bar t$
slices, where $\Sigma^{(2)} _t$ appears either as an open string 
$\lk \bar t _i <\bar t < \bar t _f \rk$ whose boundary points
represent the vortex or as an closed string $\lk \bar t _f < \bar t < \bar t _s
\rk $ in the absence of vortex flux.}
\label{fig2neu}
\end{figure}

\begin{figure}
\centerline{
\epsfysize=14.0cm
\epsffile{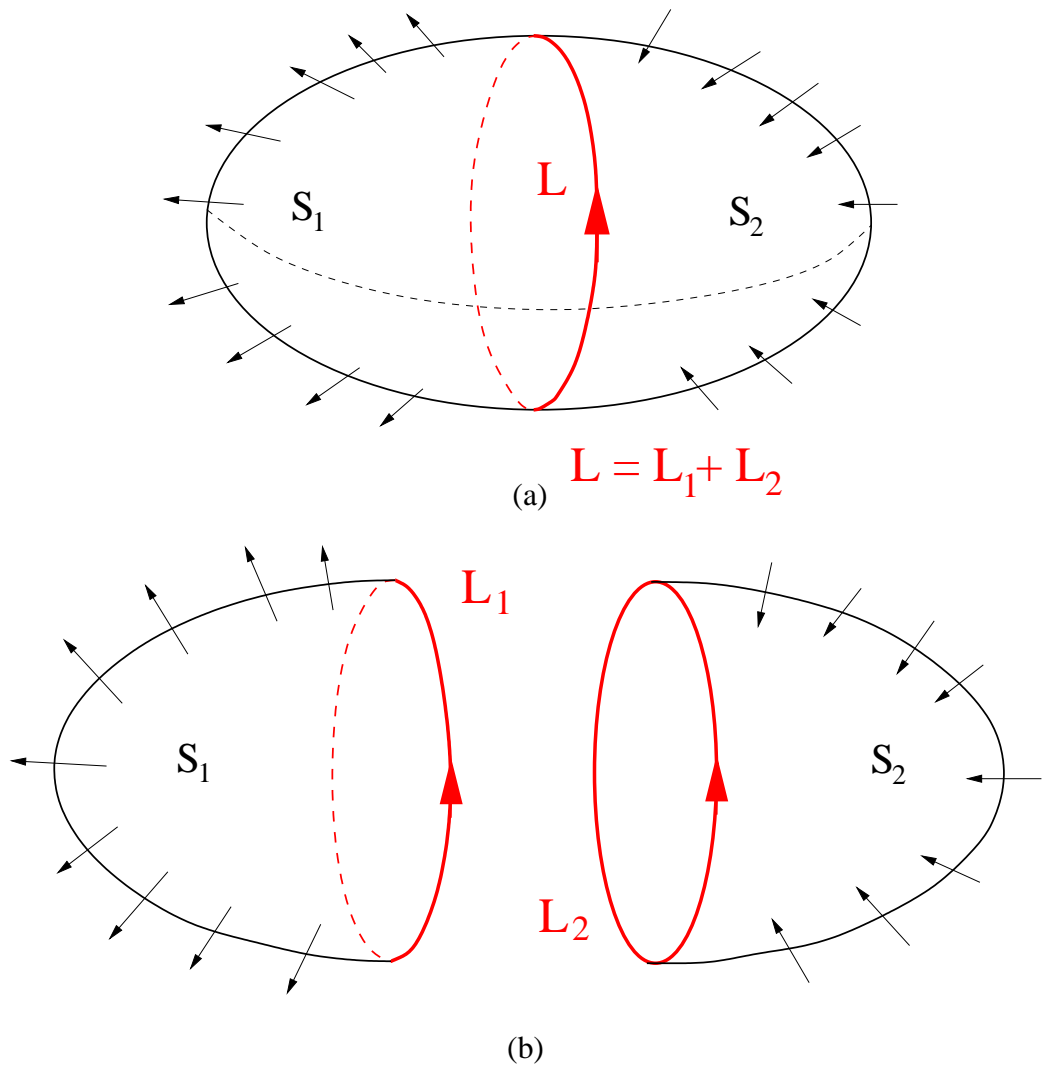}
}
\caption{(a) Non-oriented vortex sheet consisting of two oppositely oriented vortex
patches $S_1$ and $S_2$ joined by a magnetic Dirac monopole. (b) The non-oriented vortex sheet can
be considered as arising by glueing together two open oriented vortex patches
$S_1$ and $S_2$ bounded by center monopole loops 
$L_1$ and $L_2$, respectively, which add up to a Dirac monopole loop $L = L_1 +
L_2$ on the non-oriented vortex shown in (a). Note that $S_2$ can be considered as a
deformation of $S_1$ which turns the inside of $S_1$ into the outside of $S_2$.
This explains why $S_1$ and $S_2$ are bounded by identical
(center) monopole loops, although they have opposite orientation.}
\label{fig2}
\end{figure}

A generic 3-dimensional hypersurface $\Sigma$ whose boundary $\partial \Sigma$
represents a center vortex will in general be neither purely temporal
nor purely spatial. However, one can exploit the $Z(N)$ gauge freedom to split
$\Sigma$ into  purely spatial and temporal parts

\be
\label{XYZ}\Sigma = \Sigma_s \cup \Sigma_t
\ee

and, furthermore, to move the spatial part $\Sigma_s$ to such a time $t_s$ (e.g.
to $t_s
\rightarrow \pm \infty$) where no flux occurs. Let us illustrate this, for
simplicity, in $D=2+1$ dimensions where the center vortices are closed loops
$\partial \Sigma ^{(2)} = C$, which are the boundaries  of open 2-dimensional
discs $\Sigma^{(2)}$ (see fig.\ref{fig2neu}). The vortex gauge potential has support only on
this disc $\Sigma^{(2)}$. Generically this disc will take a position in space
such that $\Sigma^{(2)}$
is neither purely temporal nor spatial, i.e. $\Sigma^{(2)}\not= \Sigma^{(2)}_s
\cup \Sigma^{(2)}_t$, see fig.\ref{fig2neu}(a). By performing a $Z(N)$ gauge
transformation one can deform the area $\Sigma^{(2)}$ into the surface
$\bar\Sigma ^{(2)}$, (with $\partial \Sigma ^{(2)} = \partial \bar\Sigma ^{(2)}$) shown in
fig.\ref{fig2neu}(b), which has the shape of an open cylinder. The mantle part
$\Sigma_t^{(2)}$ of
this surface is temporal while its face $\Sigma_s ^{(2)}$ at $t=\bar t _s$ is spatial, so
that indeed $\bar\Sigma ^{(2)} = \Sigma^{(2)}_t \cup \Sigma^{(2)}_s$. 
\medskip

Obviously, the
splitting (\ref{XYZ}) of $\Sigma$ into purely temporal and spatial parts can be
achieved for both generic and non-generic center vortices\footnote{In the $D=2+1$
dimensional case considered in fig.\ref{fig2neu} a non-generic vortex would
be a planar loop in a plane perpendicular to the time axis.}. This shows that the gauge potential of a center
vortex (in singular gauge) can be chosen to satisfy the Weyl gauge $A_0 =0$ except at a
single time instant which can be chosen at will.
\medskip

In the following I will assume that the $Z(N)$ gauge has been chosen
such that eq.~(\ref{XYZ}) is satisfied
and that the spatial part $\Sigma_s$ of the hypersurface $\Sigma$ is
shifted to a time $t_s$ where no flux exists.

\no
\section{The topological charge of center vortices in terms of intersection points}
\bi

\no
The topology of gauge fields is characterised by the topological charge
(Pontryagin index)

\be
\label{G25}\nu [A] = - \frac{1}{16 \pi^2} t r \int d^4 x F_{\mu \nu} \tilde{F}_{\mu \nu}
\hk .
\ee

The field strength of an (ideal) center vortex (\ref{G11}) is given by
eq.~(\ref{F2}). Note that the flux of these vortices is indeed concentrated on the closed surface
$\partial \Sigma$ and, furthermore, the direction of the flux is determined by the
orientation of the vortex sheet $\partial \Sigma$. For definiteness, let us stick
in the following to the gauge group SU(2) where $E = 2\pi T_3, T_3 =
-\frac{i}{2} \tau_3$. The
generalization to SU(N) is straightforward. Inserting (\ref{F2}) into the
expression
for the Pontryagin index (\ref{G25}) and using $tr (EE) = -2\pi^2$ one finds
\cite{[Eng]}

\be
\label{G27}\nu [ \cA (\Sigma) ] = \frac{1}{4} I \lk \partial \Sigma^{(3)}, \partial
\Sigma^{(3)} \rk \hk ,
\ee

where

\be
\label{G28}I \lk S_1, S_2 \rk = \frac{1}{2} \int\limits_{S_1} d \sigma_{\mu \nu}
\int\limits_{S_2} d
\tilde{\sigma}'_{\mu \nu} \delta^{(4)}\lk \bar{x} (\sigma) - \bar{x} (\sigma') \rk
\ee

is the oriented intersection number of two 2-dimensional (in general open)
surfaces $S_1, S_2$ in $R^4$. Generically, two 2-dimensional surfaces intersect
in $R^4$ at isolated points. Obviously, the (self-)intersection number $I
(\partial \Sigma, \partial \Sigma)$ receives contributions only from those
points $\bar{x} (\sigma) = \bar{x} (\sigma')$, where the intersecting surface
patches give rise to four linearly independent tangent vectors (otherwise $ d
\sigma_{\mu \nu} d \tilde{\sigma}'_{\mu \nu} = 0 $). Such points are referred
to as singular points. One can distinguish two principally different types of
singular points:
\begin{enumerate}
\item transversal intersection points, for which $\bar{x} (\sigma) = \bar{x}
(\sigma')$ and $\sigma \neq \sigma' ,$ 
\item twisting points, for which $\bar{x} (\sigma) = \bar{x}
(\sigma')$ and $\sigma = \sigma'.$
\end{enumerate}
\bi

\no
Transversal intersection points arise from the intersection of two different
surface patches while twisting points occur on a single surface patch
(see sect.~5 for more details). 
Transversal intersection points yield a contribution $\pm$ 2 to the (oriented)
intersection number $ I \lk \partial \Sigma , \partial \Sigma \rk $, 
where the sign depends on the relative orientation of the two
intersecting surface pieces. (One should note that each transversal
intersection point actually contributes twice to the self-intersection number).
For orientable (and oriented) closed surfaces twisting points can be turned into
(a smaller number of) transversal intersection points by surface deformation. In
this sense for closed surfaces in $D = 4$ the number of transversal intersection
points is even, so that the topological charge of orientable center vortex
surfaces is indeed
integer-valued. However, for closed oriented surfaces the oriented
self-intersection number eq.~(\ref{G28})
vanishes. Hence, vortices
carrying non-zero topological charge are given by non-oriented surfaces, where
the orientation defines the direction of the flux. Non-oriented vortex surfaces 
consist of open oriented surface patches $S_i \lk
i= 1,2,...\rk$.
The expression for the field strength eq.~(\ref{F2}) remains valid for open
surface patches and hence also for non-oriented vortex sheets.
\medskip

Using eq.~(\ref{F2}) for an oriented open surface patch $S$ one finds upon
using Stokes' theorem that the monopole current

\be
\label{G29} 
j_\mu \lk \partial S ,x \rk = \partial_\nu \tilde{\cal F}_{\mu \nu}
\lk S,x \rk = E \oint\limits_{\partial S} d \bar{x}_\mu \delta^{(4)}\lk
x-\bar{x}\rk 
\ee

flows at the boundary $\partial S$ of the vortex patch $S$. The magnetic charge
$m$ of the monopole is obtained by integrating the current
(\ref{G29}) over the 3-dimensional volume $\Sigma$ dual to the monopole trajectory $\partial S$.
This yields

\be
\label{G30}
m & = & E \int\limits_\Sigma d\Sigma_\mu \oint\limits_{\partial S} \delta x_ \mu \delta^{(4)} \lk
\bar{x} (\Sigma) -x \rk = E I \lk \Sigma , \partial S \rk , 
\ee

where we have used the definition of the intersection number $I \lk \Sigma ,
\partial S \rk$ (\ref{G28}), which, by the choice of $\Sigma$, is equal here to one. Hence
the magnetic charge of the monopole flowing at the boundary of an oriented
vortex patch is given by $E$ (\ref{G6}). In view of eq.~(\ref{G6}) we call this monopole a center
monopole. 
\medskip

Consider a non-oriented center vortex sheet $\partial \Sigma$ which
consists of two oriented patches $S_1$ and $S_2$ (see fig. \ref{fig2}). Since $S_1$ and
$S_2$ are oppositely oriented and form together a closed surface, they possess the
same boundary $\partial S_1 = \partial S_2$. 
Hence the two magnetic monopole loops flowing along $\partial S_1$ and
$\partial S_2$ , respectively, (each with charge E) add up coherently to form the
trajectory of a magnetic monopole with twice the charge of a center monopole. 
In view of $e^{-2E} = 1$ this monopole is a Dirac monopole whose magnetic charge
is in accord with Dirac's quantization condition. Hence a magnetic (Dirac)
monopole with magnetic charge $2E$ flows at the boundary between two oppositely
oriented patches of a (globally non-oriented) center vortex. Thus non-oriented
vortex sheets necessarily carry magnetic (Dirac) monopoles. 
\medskip

Indeed, a non-oriented vortex surface $\partial \bar{\Sigma}$ can be interpreted
as an oriented vortex surface $\partial \Sigma$ covered with an (open)
oppositely oriented Dirac sheet $S_D, \partial \bar{\Sigma} = \partial \Sigma +
S_D$. The Dirac sheet $S_D$ carries twice the flux of a center vortex, and thus, if it
is oppositely oriented to the vortex surface $\partial \Sigma$, it will reverse
the orientation of the latter. Furthermore the boundary of the Dirac sheet represents the world line of a Dirac
monopole (here with magnetic charge $2E$). Thus non-oriented closed magnetic vortex sheets consist of oriented surface
patches joined by magnetic monopole loop currents.
The magnetic monopole currents change the orientation of the center vortices and 
are thus absolutely necessary for a non-vanishing topological charge of center
vortices, see ref. \cite{[Eng]} for more details. 
\medskip

The orientation of - and thus the magnetic monopole loops on - the center vortices
are, however, not relevant for the confining properties of the latter, which are entirely
determined by their linking properties with the Wilson loop, see eq.~(\ref{G1}).
This can be seen by noticing that non-oriented vortex surfaces differ from
oriented ones by open Dirac sheets which do not affect Wilson loops.
\medskip

A final comment is in order: Eq.~(\ref{F2}) still applies to the field
strength of non-oriented vortex sheets $\partial \bar{\Sigma}$. However, the
field strength of non-oriented vortex sheets cannot be generated from a
globally defined Abelian gauge potential. Indeed the flux of the open Dirac
sheet (converting the oriented vortex into the non-oriented one) cannot be 
represented as the curvature of an Abelian gauge
potential. The monopole current at the boundary of the Dirac sheet violates the
Abelian Bianchi identity $\partial_\mu \tilde{{\cal F}}_{\mu \nu} = 0$, which holds for
any field strength constructed from a regular Abelian gauge potential.
Furthermore, though eq.~(\ref{F2}) applies also to non-oriented vortex sheets and
includes in particular the flux of the involved Dirac string it does not contain
the magnetic field of the monopole itself. This might at first sight seem
disturbing but in fact is a very welcome feature mimicking the non-Abelian
nature of magnetic monopoles in Yang-Mills theory. This can be understood as
follows. In Yang-Mills
theory the magnetic monopoles arise as gauge fixing defects in the so-called
Abelian gauges where one fixes only the coset $G/H$ leaving the Cartan subgroup
$H$ unfixed. To be more precise if $V \in G/H$ 
is the gauge transformation  $A \rightarrow A^V = V A V^\dagger +
V\partial V^\dagger$ necessary to bring the gauge field into the desired Abelian
gauge, 
the magnetic monopoles arise in the Abelian part of the induced gauge potential
$\cA = V \partial V^\dagger $. As shown in ref. \cite{[Lange]}, (see also ref.
\cite{[Reinh]}) the Abelian field strength 
$ f^n_{\mu \nu} = \partial_\mu \cA^n_\nu - \partial_\nu \cA^n_\mu $ 
($\cA^n$ denotes the Abelian part of $\cA$) 
contains the 
magnetic monopole with a Dirac string attached to it. Now the
crucial point is that the Abelian part of the commutator $\left[\cA_\mu,
\cA_\nu \right]^n$ contains the same, but oppositely directed monopole field (i.e. the
corresponding anti-monopole field), however, without
the Dirac string, so that in (the Abelian part of) the total field strength

\be
\label{G36}
F_{\mu \nu} \lke V \partial V^\dagger \rke^n & = & f^n_{\mu \nu} + \left[
\cA_\mu, \cA_\nu \right]^n\nonumber\\
& = & \lk V \lk \partial_\mu \partial_\nu - \partial_\nu \partial \mu \rk
V^\dagger \rk^n
\ee

the monopole does not show up while the Dirac string is left. In fact the 
total field strength vanishes (as it
should for a pure gauge) except at the  Dirac sheet, where $V$ is singular.
Thus, the total field strength, eq.~(\ref{G36}), contains only the flux of the Dirac
sheet. 
\medskip

The conclusion from the above considerations is that Dirac sheets
(without monopole fields) naturally arise in the total field strength of
non-Abelian gauge fields and are induced by the singular gauge transformations
necessary to implement the Abelian gauges. This observation is consistent with
the picture advocated in ref. \cite{[Debbio]}:
As function of space-time-coordinates the non-Abelian gauge field changes
smoothly and gradually in color space. Abelian
gauge fixing rotates the gauge field as much as possible into the Abelian
direction. The flux is, of course, not changed during the gauge fixing and hence
still smooth. Upon Abelian projection the regions interpolating between positive
and negative Abelian flux directions appear as magnetic monopoles. 
The monopoles are an artifact of the Abelian projection and do not show up in the full
non-Abelian flux. The full non-Abelian field strength can, however, contain
string like fluxes: Dirac sheets (strings) or center vortices, which may be
interpreted as half Dirac sheets.

\no
\section{Topology of generic center vortices}
\subsection{The topological charge of generic center vortices}
\bi

As is well known the topological density can be expressed as a 4-dimensional
divergence

\be
\label{G68}
- \frac {1}{16 \pi^2} tr F_{\mu \nu} \tilde{F}_{\mu \nu} = \partial_\mu K_\mu
\ee

where 

\be
\label{G69}
K_\mu = - \frac{1}{8\pi ^2} \epsilon_{\mu \alpha \beta \gamma}tr \lke
A_\alpha \partial_\beta A_\gamma + \frac{2}{3} A_\alpha A_\beta A_\gamma \rke 
\ee

is the topological current. For simplicity let us choose the singular gauge
(\ref{G11}), (\ref{G22}) where the gauge potential of a (closed) center vortex sheet 
$\partial \Sigma$
has support on an (open) 3-dimensional volume $\Sigma^{(3)}$ in the
4-dimensional space-time manifold. As shown in sect.2  $\Sigma^{(3)}$ can be
chosen to be the sum of a purely spatial part $\Sigma_s ^{(3)}$ (existing only at
a single time instant), and a purely temporal part $\Sigma_t ^{(3)}$, which at a
fixed time $t$ represents a 2-dimensional open (spatial) disc  $\Sigma_s
^{(2)}(t)$. Since
$\Sigma_s ^{(3)}$ and $\Sigma_s ^{(2)}$ are both localised objects in ordinary
3-space $M = \mathbbm{R}^3$ the topological current $K_\mu \lk \vec{x} ,t \rk $ is a 
localised function of the
spatial coordinates which vanishes for $|\vec x |\rightarrow \infty$.
Hence by applying Gauss' law we obtain

\be
\label{G73}
\int \limits_{M} d^3 x \partial_i K_i (x) = \oint \limits_{\partial M}
d\tilde{\sigma}_i (x) K_i (x) = 0.
\ee

This relation is not spoiled for non-oriented vortex sheets in the presence of
magnetic monopole loops on the vortex sheets. For a generic center vortex sheet,
at a fixed time, the monopole loop represents the positions of a monopole and an
anti-monopole. At large distance the magnetic monopole -anti-monopole pair
produces a magnetic dipole field, which drops off sufficiently rapid not to
contribute to the surface term (\ref{G73}).
\medskip

With (\ref{G73}) the topological charge (\ref{G25}) becomes 

\be
\label{G74}\nu = \int d^4 x \partial _0 K_0 = \int dx_0 \partial_0 S_{CS}[A] \lk x_0\rk
\ee

where 

\be
\label{G75}S_{CS} [A] \lk x_0 \rk = \int d^3 x K_0 [A] \lk x_0 ,\vec{x}\rk 
\ee

is the Chern-Simons action. 
\medskip

If we assume that our space-time manifold extends over a finite time interval,
saying from $t_i$ to $t_f$, one would naively expect that the topological charge
becomes 

\be
\label{G76}\nu = S_{CS} [A] \lk t_f\rk - S_{CS}[A] (t_i).
\ee

This is the well-known expression for the topological charge in Weyl gauge $A_0
= 0$ , in which gauge fields are considered to be smooth. However, although the gauge potential of center vortices can be chosen to
satisfy the Weyl gauge (except at a single time instant, see sect.2), depending
on the gauge configurations under consideration,
their
Chern-Simons action need not to be differentiable in the whole time interval, but
may jump at intermediate time instants, say $\bar{t}_k , k= 1,2, ...$, which we will
refer to as $singular$ $time$ $instants$\footnote{The idealised center vortices
which I consider in the present work can be thought of as arising by center
projection from initally smooth gauge field configurations. It is the center projection
which introduces the singularities.}. Each of these jumps contributes to the
topological charge 

\be
\label{G77}\Delta \nu = \Delta S_{CS}^{(k)} = \lim \limits _{\varepsilon \rightarrow 0} \lk
S_{CS}\lk \bar{t}_k +\varepsilon \rk - S_{CS}\lk \bar{t}_k - \varepsilon\rk \rk 
\ee

These discontinuities in the Chern-Simons action will be investigated for center
vortices in more detail further below, see sects. 4.2 and 5.

\no
\subsection{The Chern-Simon action of center vortices}
\bi

The ideal center vortices considered above live in the Cartan subgroup.
Consequently for these configurations the Chern-Simons action reduces to the
Abelian one 

\be
\label{G78}
S_{CS} [A](t) = \frac {1}{8 \pi^2} \int d^3 x \mathbf{A} \cdot \mathbf{B}
\ee

where $\mathbf{B} = \vec{\partial} \times \mathbf{A}$ denotes the magnetic field of the
vortex. For a generic center vortex describing a magnetic flux loop $C(t)$
which evolves in time, the magnetic field is given by (\ref{ss}).
Inserting this expression into eq.~(\ref{G78}) and using the vortex gauge
potential in the singular gauge (\ref{G22}) for $A_i (x)$ we 
immediately obtain for the Chern-Simon action of a generic center vortex $C(t) =
\partial \Sigma _s ^{(2)} (t)$ evolving in time

\be
\label{CS}S_{CS} \lke A \lke \Sigma_s ^{(2)} \lk t \rk \rke \rke = \frac{1}{4} I \lk
\Sigma_s ^{(2)} \lk t \rk , C \rk 
\ee

where

\be
I \lk \Sigma , C \rk = \int\limits_{\Sigma} d^2 \tilde{\sigma} _i
\oint\limits_{C} dx_i \delta^{(3)}\lk x - \bar x (\sigma) \rk
\ee

is the intersection number between the open 2-dimensional surface $\Sigma$ and
the loop $C$ in $\mathbbm{R}^3$.
This intersection number $I \lk \Sigma , C \rk$ equals the Gaussian linking
number $L \lk \partial \Sigma , C \rk $ between the boundary $\partial \Sigma$
of the surface $\Sigma$ and the loop $C$, which is defined for two closed loops
in $\mathbbm{R}^3$ by 

\be
\label{G82}
L \lk C_1 ,C_2 \rk & = & -\varepsilon _{i j k} \oint \limits_{C_1} dx_i \oint
\limits_{C_2} d x' _j \partial_k D^{(3)}\lk x -x' \rk \nonumber\\
& = & \frac{1}{4\pi} \oint \limits_{C_1} dx_i \oint\limits_{C_2} d x'_j
\varepsilon_{i j k} \frac {x_{k} - x'_{k}}{\left | \vec{x} -\vec{x'}\right |^3}.
\ee

(Note that for $\vec{x} \rightarrow \vec{x'}$ the integrand (including the
measure) does not diverge but vanishes). With $I \lk \Sigma , C \rk = L
\lk \partial \Sigma , C \rk$ we obtain from eq.~(\ref{CS}) for the Chern-Simons 
action of a generic center vortex loop $C(t)$.

\be
\label{G80}
S_{CS}\lke A \lke C(t)\rke \rke = \frac {1}{4} W (C) 
\ee

where $W (C)$ is the writhing number of the closed (time-dependent) vortex loop
$C (t)$, which is defined as the coincidence limit

\be
\label{G81}
W (C) = L (C,C)
\ee

of the Gaussian linking number. 
In view of equations (\ref{G74}) and (\ref{G80}) we find that the topological charge of
an (idealised) generic center vortex, representing a time dependent closed magnetic
flux loop $C(t)$, is given by 

\be
\label{G83}\nu = \frac {1}{4} \int dt \partial_t W \lk C (t) \rk . 
\ee 

Let us again assume that our space-time manifold has a finite extension in time
direction, i.e. extends from an initial time $t_i$ to a final time $t_f$.
According to eq.~(\ref{G83}) the topological charge is given by the changes of the
writhing number during the evolution from the initial time $t = t_i$ to the final
time $t = t_f$. If $W (t)$ changes continuously during the whole time
evolution the topological charge is given by 

\be
\label{G84}\nu = \frac{1}{4} \lk W (t_f ) - W (t_i )\rk,
\ee  

so we need not to consider any smooth change of $W (t)$ during the time
evolution in the calculation of the topological charge, which then solely depends only on the
initial and final values of $W (t)$.
However, the writhing number $W (t)$ may change in a discontinuous way, e.g. when two line
segments of the vortex loop intersect (see below).
These discontinuous changes (due to singular changes in the vortex loop), which also
contribute to eq.~(\ref{G83}), are left out in eq.~(\ref{G84}).           
\medskip

If we denote by $\bar{t}_k , k= 1,2,..; t_i < \bar{t}_k < t_f$ the
intermediate time instants where $W(t)$ jumps by a finite amount 

\be
\label{G85}\Delta W \lk \bar{t}_k \rk = \lim \limits_{\varepsilon \rightarrow 0} \lke W \lk
\bar{t}_k + \varepsilon \rk - W \lk t_k - \varepsilon \rk \rke
\ee

the complete expression for the topological charge for a generic center
vortex is given by

\be
\label{G86}\nu = \frac{1}{4} \lke W \lk t_f \rk + \sum \limits_k \Delta W \lk \bar{t}_k \rk
- W \lk t_i \rk \rke
\ee

This relation will be illustrated in sect. 5 by means of an example.

\no
\subsection{The self-linking number, the writhing number and the twist }
\bi

Since the topological charge of generic center vortices is completely
determined 
by the temporal changes of the writhing number, see eq.~(\ref{G83}), let us recall some
of its properties.
The writhing number is defined as coincidence limit of the Gaussian linking
number, see eq.~(\ref{G81}). However, unlike the Gaussian linking number $L(C_1 ,C_2)$ of two
closed loops $C_1 ,C_2$ the writhing number $W (C)$ is not a topological
invariant, but depends on the precise shape of the loop C. In particular it is not
integer but real-valued. 
\medskip

For a single loop $C$ one can define the integer-valued, topologically
conserved self-linking number as the Gaussian linking number between the loop $C$ under
consideration and a loop $C'$ arising by displacing the original loop $C$
by an infinitesimal amount $\varepsilon$ parallel to a unit vector $\hat{n}\lk
\vec{x},t\rk$, which has to be always orthogonal to the loop $C$. This displacement
introduces a so-called framing of the loop by which the latter becomes a
band or a ribbon bounded by C and C' (c.f.~fig.~\ref{fig7}, below). 
The self-linking number defined by 

\be
\label{G87}
S L \lk C, \hat{n} \rk = \lim \limits _{\varepsilon \rightarrow 0} L \lk
C, C' = C + \varepsilon \hat{n}\rk
\ee

is frame dependent but integer-valued (since it represents the linking number of
$C$ and $C'$). Taking the coincidence limit $\varepsilon \rightarrow 0$
one finds 

\be
\label{G88} S L\lk C, \hat{n}\rk = W (C) + T \lk C,\hat{n}\rk
\ee

where $W (C)$ is the writhing number defined above by eqs.~(\ref{G81}) and
(\ref{G82}) and

\be
\label{G89}T (C, \hat{n}) = \frac {1}{2\pi} \oint \limits_C ds \frac
{\vec{r} (s)} {\left | \vec{r} (s) \right |} \lk \dot{\hat{n}} (s) \times \hat{n}(s)\rk
\ee

is the so-called ``twist''. The latter represents the integrated torsion of the ribbon
bounded by C and $C'$. Furthermore $\vec{r}(s)$ denotes a parametrisation of the
loop $C$. The twist $T (C, \hat n)$ can take any real value and also depends on the framing.
Let us emphasise, while the self-linking number $S L (C,n)$ and the twist $T
\lk C, \hat{n} \rk$ are framing dependent, their difference, the writhing number $W
(C)$, is independent of the chosen framing. Furthermore $W (C)$ and $T
(C,\hat{n})$ can take any real value, but their sum $S L (C, n)$ is integer
valued though framing dependent. Since the self-linking number is
integer-valued and $W (C)$ is frame independent, eq.~(\ref{G88}) implies that
changing the framing can change the twist (torsion) only by an integer. 
\medskip

\begin{figure}
\centerline{
\epsfysize=17.0cm
\epsffile{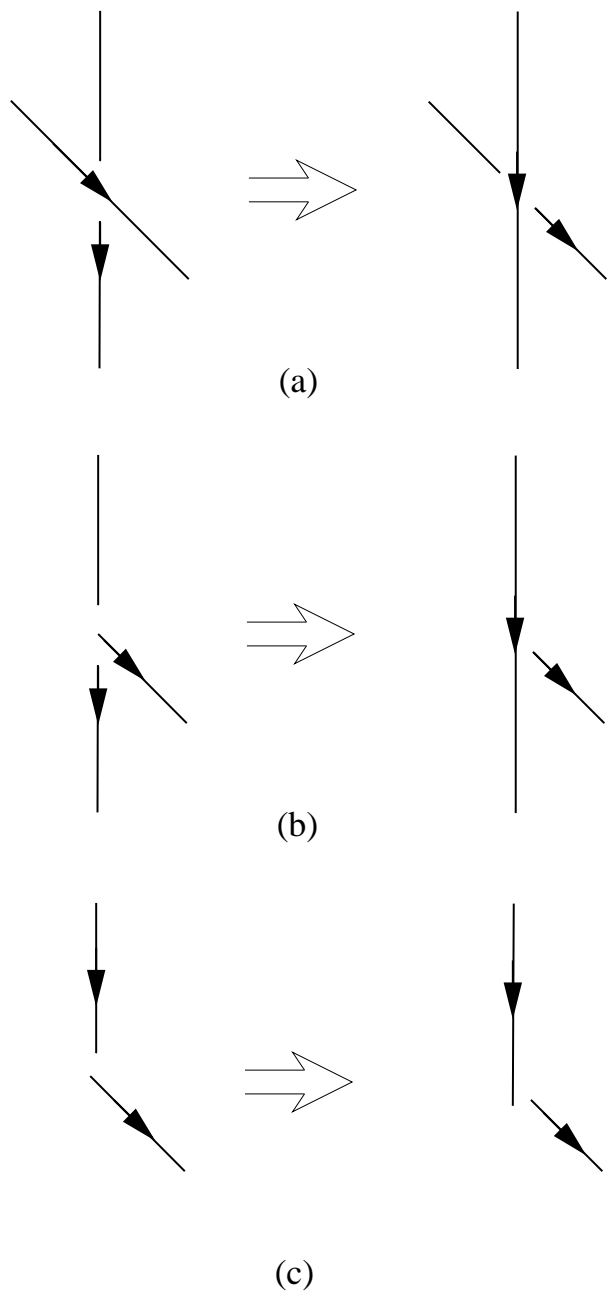}
}
\caption{Crossing of (a) two (full) line segments, 
(b) a full and a half-line and
(c) two half-line segments of a closed loop changing the writhing of the loop by
$\Delta W = 2, \Delta W = 1$ and $\Delta W = \frac{1}{2}$, respectively.}
\label{fig3}
\end{figure}

As already mentioned above, the writhing number $W (C)$ is a
continuous function of the shape of the curve $C$ but it is not a topological
invariant. From its definition, eqs.~(\ref{G81}), 
(\ref{G82}), immediately follows that $W (C)$ vanishes for planar loops $C$, for which the integrand in
eq.~(\ref{G82}) vanishes identically. More generally, $W (C)$ vanishes for
curves possessing a symmetry plane. In this sense the writhing number measures the
left-right asymmetry of a loop, i.e. its chirality. 
Furthermore, $W (C)$ suffers discontinuities when two segments of the loop $C$ pass through
each other: When two line segments cross (i.e. intersect) (see fig. \ref{fig3} (a)) $W (C)$
changes by $\pm 2$, where the sign depends on the relative orientation of the crossing line
segments. Furthermore, what is not commonly known and will be demonstrated later on when we consider
explicit examples, when two half-line segments cross, see fig. \ref{fig3} (c), which can
be interpreted as a twist (see sect. \ref{fig4}), the writhing number changes by $\Delta W = 2 \cdot
\frac{1}{2} \cdot \frac{1}{2} = \frac {1}{2}$. Analogously, when one half-line
crosses with a full line (see fig. \ref{fig3} (b)) the writhing number changes by $\Delta W = 2 \cdot 1
\cdot \frac{1}{2} = 1$. All these properties of the writhing number will
explicitly show up in the next section where we analyse the distribution of
topological charge of a specific center vortex configuration. 

\no
\section{Analysis of a lattice center vortex configuration in continuum
Yang-Mills theory}
\subsection{The topological charge of lattice center vortices}
\bi

On the lattice, idealised center vortices are co-closed hypersurfaces built up from
plaquettes which are equal to a non-trivial center element. In addition these
vortex surfaces have to be endowed with an orientation, which defines the direction of the vortex flux.
The topological charges of such vortex hypersurfaces can in view of eq.~(\ref{G27}), in
principle,  be easily determined by measuring the (oriented) intersection
number: The topological charge at a lattice site $x$ is obtained by counting all
pairs of dual plaquettes\footnote{The dual $\tilde{P}_{\mu \nu}$ of a plaquette $P_{\mu \nu}$ is defined
as usual by $\tilde{P}_{\mu \nu} = \frac{1}{2}  \epsilon_{\mu \nu \alpha \beta}
P_{\alpha \beta}$.} meeting at the lattice site considered, i.e. one
finds for the topological charge $\nu_x$ of lattice site $x$ \cite{[Eng2]}

\be
\label{G90} \nu_x = \frac{1}{32} N_x ,
\ee

where $N_x$ is the number of pairs of dual plaquettes meeting at site
$x$. As discussed above, we can basically
distinguish two types of singular points of vortex sheets contributing to the
topological charge: transversal intersection points and twisting points. On the
lattice a transversal intersection point
corresponds to a site $x$ where two mutually dual planar vortex patches, each
consisting of four plaquettes meeting at $x$ , intersect. Thus $N_x = 4\times 4 =
16$ pairs of dual vortex plaquettes meet at a transversal intersection point,
yielding $\nu_x = \frac{1}{2}$, in agreement with the continuum result (see sect.
3 and ref. \cite{[Eng]}). A twisting point at lattice site $x$
consists of a single (non-planar) surface segment twisting around $x$ in such a way
to produce a non-zero contribution to the intersection number. Such a surface segment is 
 composed by plaquettes which meet at $x$ and which
can be connected by proceeding along plaquettes which share a link \cite
{[Eng2]}\footnote{In ref. \cite{[Eng2]} twisting points were referred to
as writhing points, since in a pictorial sense, the vortex surface writhes
similar to the winding of a screw. However, from a mathematical point of view it
makes sound sense to refer to them as twisting
points. This is because, these points can change both the writhing number and the
twist, while the intersection points change only the writhing number but never
the twist, as we will illustrate later.}. For twisting
points $N_x < 16$ holds.
Obviously on the lattice the topological charge at a site  $x$, see 
eq.~(\ref{G90}), can take only integer multiples of $\frac{1}{32}$ since $N_x$ is integer valued. 
\medskip

To illustrate the various singular vortex points, let us consider as an example
the
center vortex configuration shown in fig. \ref{fig4}. \cite{[Eng2]}, which could
arise in a lattice
simulation after center projection, ref. \cite{[Ber]}, or in a random vortex
model \cite{[Eng3]}. This vortex surface is orientable.
For simplicity let us also assume that its flux is indeed oriented, so that
there are no magnetic monopole loops on this vortex sheet. The total topological
charge of this configuration then vanishes, $\nu =0$. Nevertheless this
configuration has various spots of non-zero topological charge \cite{[Eng2]}. 
There is a transversal intersection point at the intermediate time\footnote{We
quote here the time $t = n_0 a$ in (integer) units $n_0$ of lattice spacing
$a$.} $n_0 =2$
contributing $\frac{1}{2}$ to the topological charge $\nu$. At this time there
are also two twisting points at the front and back edges of the configuration
each contributing $- \frac{1}{8}$ to $\nu$. Further twisting points occur at the initial $(n_0 =1)$ and
final $(n_0 =3)$ times, each contributing $-\frac{1}{8}$ to $\nu$ so that the
total topological charge vanishes $(\nu = 0)$ for this vortex configuration,
indeed. 

\no
\subsection{The writhing number of center vortex loops}
\bi

Let us now interpret the same configuration as a time dependent vortex loop
in ordinary 3-dimensional space (like in a movie-show) eliminating lattice
artifacts
due to the use of a discretised time.  As illustrated in fig. \ref{fig5} purely spatial vortex patches are lattice artifacts.
They represent the discrete time step approximation to continuously evolving (in
time) vortex loops. Fig. \ref{fig6} shows the time-evolution of a closed magnetic vortex
loop in ordinary 3-dimensional space which on the 4-dimensional lattice gives rise to the
configuration shown in fig. \ref{fig4}. For simplicity we have kept the cubistic
representation in $D = 3$ space, so that the loop consist of straight
line segments $C_i$ defined in fig. \ref{fig11}. At an initial time $t = \bar{t}_i$ an infinitesimal
closed vortex loop is generated which then growths (i.e. continuously evolves) up to a time
$t =\bar{t}_1$. Then the loop segment $C_5$ moves towards, and at time $t =
\bar{t}_2$ crosses the segment $C_1$, and continues to move up to a time $t =
\bar{t}_3$. After this time the loop decreases continuously and at the fixed time
$t = \bar{t}_f$ shrinks to a point. In the following we calculate the
topological charge of this time dependent vortex configuration by applying
eq.~(\ref{G86}).
\medskip

For simplicity let us choose $t_i < \bar{t}_i$ (the birth of
the vortex loop) and $t_f > \bar{t}_f$ (the death of the vortex loop). Then 

\be
\label{G91}W (t_i) = W (t_f) = 0
\ee
since there are no vortex loops at the initial and final times. In the above field 
configuration shown in fig. \ref{fig6} there are discontinuous changes of the vortex
loop, and accordingly of the writhing
number, at the creation (birth) of the vortex loop at $t = \bar{t}_i$, at the
intermediate time $\bar{t}_2$,  where two line segments cross and two lines turn by 180
degrees, and at the annihilation (death) of the vortex loop at $t = \bar{t}_f$. Hence the
topological charge of this configuration is given by (assuming $t_i < \bar{t}_f$
and $t_f > \bar{t}_f$) 

\be
\label{G92}\nu = \frac{1}{4} \lke \Delta W \lk \bar{t}_f \rk + \Delta W \lk \bar{t}_2 \rk +
\Delta W \lk \bar{t}_i\rk \rke
\ee 

For calculational simplicity let us assume that from its creation at
$\bar{t}_i$ until the time $\bar{t}_1$ the vortex loop does not change its
shape, but merely scales in size. The same will be assumed for the vortex evolution
from $\bar{t}_3$ until its annihilation at $\bar{t}_f$. Then the change of the
writhing number at vortex creation and at annihilation, respectively, is given by
(with $ W \lk \bar{t}_i - \epsilon \rk = 0, W \lk \bar{t}_i + \epsilon \rk = W \lk 
\bar{t}_1 \rk , W \lk \bar{t}_f + \epsilon \rk = 0 , W \lk \bar{t}_f - \epsilon \rk = W \lk \bar{t}_3
\rk $)

\be
\label{G93}\Delta W (\bar{t}_i) = W (\bar{t}_1) , \hspace{0.5cm} \Delta W (\bar{t}_f) = - W
(\bar{t}_3),
\ee

so that we obtain for the topological charge (\ref{G92})

\be
\label{nu}\nu = \frac{1}{4} \lke W (\bar{t}_1) + \Delta W (\bar{t}_2) - W
(\bar{t}_3) \rke .
\ee

The writhing numbers $W (\bar{t}_1) ,  W (\bar{t}_3)$ are explicitly evaluated
in Appendix A. One finds

\be
\label{G94}W (\bar{t}_1) = \frac{1}{2} , \hspace{0.5cm} W (\bar{t}_3) = - \frac{1}{2}
\ee

As already stated above, a further singular change of the vortex loop shown in
fig. \ref{fig6} occurs at the intermediate time $t=\bar{t}_2$. Since the writhing number
$W (C)$ is a unique function of the shape of the loop $C$ (independent of any
framing) the changes of $W$ at $t =\bar{t}$ are obviously given by

\be
\label{zz} \Delta W (\bar{t}_2 ) = W (\bar{t}_3 ) - W (\bar{t}_1 ) = -1
\ee

Let us now investigate how this contribution arises from the various singular
points at $t = \bar{t}_2$.
At this time the line segment $C_5$
intersects (crosses) the line segment $C_1$
(see fig. \ref{fig6}). The crossing of these two line segments at $t = \bar{t}_2 , \lk
x,y,z \rk = \lk 0,0,0 \rk $ corresponds in D=4 to the transversal intersection point
shown in fig. \ref{fig4} at $n_0 =2$. In fact, in Appendix A it is shown that the
crossing of the line segments
$C_5 $ and $C_1$ gives rise to a change in the writhing number of 

\be
\label{G97}\Delta W \lk \bar{t}_2 \rk ^{(i)} = -2
\ee

which in view of eq.~(\ref{G86}) is in accord with the finding \cite{[Eng]} that a
transversal intersection point contributes $\Delta \nu = \pm \frac{1}{2}$ to the
topological charge.
Furthermore, when the loop segment $C_5$ crosses $C_1$ the paths $C_4$ and 
$C_6$ reverse
their directions,  which can be interpreted as twisting $C_4$ and $C_6$ by an
angle $\pi$ around the x-axis at $x = \pm a $ and at $t = \bar{t}_2 $, see
figs. \ref{fig6}, \ref{fig11}. In the $D=4$ dimensional lattice realization of the present center vortex shown in
fig. \ref{fig4} these twistings of the vortex loop segments (in $D=3$) by angle $\pi$
correspond to the two twisting points at $n_0 =2$ at the front and back edges of
the configuration. Thus we observe that twisting points of 2-dimensional vortex sheets in $D=4$  
may manifest themselves as rotations of loop segments (here by an angle $\pi$) in $D=3$. Hence we will refer to the
turning points $\lk \pm a,0,0 \rk$ of the rotations at $t = \bar{t}_2$ also as
twisting points. 
\medskip
 
As shown in Appendix A the two twisting (turning) points $\lk \pm a,0,0 \rk$ at
$t = \bar{t}_2$ both change the writhing number
by

\be
\label{G98}\Delta W \lk \bar{t}_2 \rk ^t \lk \pm a \rk = \frac{1}{2} 
\ee

and hence contribute $\Delta \nu = \frac{1}{8} $ to the topological charge,
again in agreement with the analysis of $\nu$ in $D=4$.
\medskip

The change of the writhing number by $\Delta W = \frac{1}{2}$ by the twisting
points at $\lk \pm a,0,0 \rk$ can be easily understood by noticing that the
twisting points can be interpreted as the crossing of two half-lines (see fig.
\ref{fig3} (c)) which changes the writhing number by only 
$\frac{1}{2} \cdot \frac{1}{2} = \frac{1}{4} $ 
of the change of $W \lk C(t)\rk $ at a true intersection point of two
(full) lines $\lk \Delta W^i = \frac{1}{2} \rk $ so that 

\be
\label{G100}
\Delta W ^t = \frac{1}{2} \cdot \frac{1}{2} \Delta W^i = \frac{1}{8}.
\ee

Thus at the time $t=\bar{t}_2$ the total change in writhing number is given by

\be
\begin{array}{lllll}
\Delta W \lk \bar{t}_2 \rk & =  \Delta W^i \lk \bar{t}_2 \rk 
& + W^t (\bar{t}_2 ,a) & +  \Delta W^t ( \bar{t}_2 ,-a) & \\
& = -2  & + \frac{1}{2} & + \frac{1}{2} & = -1
\end{array}
\label{G101}
\ee

This is in agreement with our previous result found in eq.~(\ref{zz}).
Hence one finds for the total change in the writhing number during the whole
time-evolution of the vortex

\be
\label{G102}
\Delta W & = & W \lk \bar{t}_1 \rk + \Delta W \lk \bar{t}_2 \rk - 
W \lk \bar{t}_3 \rk \nonumber \\
& = & \frac{1}{2} - 1 + \frac{1}{2} = 0
\ee

which corresponds to vanishing Pontryagin index.
\medskip

In view of the relation (\ref{G86}) the above results are in complete agreement with the
contributions of the various singular points to the Pontryagin index obtained in $D=4$ from the previous analysis of
the intersection number of the vortex sheet shown in fig. \ref{fig4}.
\medskip

\no
\subsection{The twist of the center vortex loops}
\bi

Let us also consider the twist or torsion (\ref{G89}) of these vortex loops.
For this purpose we have to introduce a framing of these loops, see fig.
\ref{fig7}. The
twist is evaluated in Appendix B. One finds

\be
\label{G95}
T (\bar{t}_1) = - \frac{1}{2} , \hspace{0.5cm} T (\bar{t}_3) = \frac{1}{2}.
\ee

This result for the twist can be also immediately read off from the framing:
Consider for example the framing of the vortex loop at 
$t = \bar{t}_3 $ shown in fig. \ref{fig7}(b). As one moves along the loop, 
the ribbon of the framing does not twist along $C_1$ and $C_2$. 
Along the path $C_3$ the ribbon twists by an angle $-\frac{\pi }{2} $, which,
in view of eq.~(\ref{G89}), results in a contribution 
$\lk - \frac{1}{4} \rk $ to the torsion $T(C)$. Along
$C_4$ the ribbon accumulates a twist (angle) of $\frac{\pi}{2}$ i.e. a torsion
$T(C_4 ) =\frac{1}{4}$. Furthermore the ribbon does not twist along $C_5$, 
but accumulates a torsion of $\frac{1}{4}$ along $C_6$ and also along 
$C_7$ while the ribbon does not twist along
$C_8$. Therefore we obtain for the total twist

\be
\label{G96}
\begin{array}{llllll}
T \lk C \lk \bar{t}_3 \rk \rk & = T \lk C_3 \rk & +  \lk C_4 \rk & +  T \lk C_6
\rk  & + T \lk C_7 \rk & \\
& = -\frac{1}{4} & + \frac{1}{4} & + \frac{1}{4} & + \frac{1}{4} & =
\frac{1}{2}.
\end{array}
\ee

This result agrees with that of the explicit calculation given in Appendix B.
By the identity (\ref{G88}) the results for $W$ and $T$ imply that the self
-linking number of the framed curve $C (\bar{t}_3)$ shown in fig. \ref{fig7} (b) vanishes. 
\medskip

Let us now consider the change of the twist at $t = \bar{t}_2$. Obviously, 
assuming the framings adopted above, which are shown in fig. \ref{fig7} (a), (b), the
total change of the twist at $t = \bar{t}_2$ is given by 

\be
\Delta T \lk \bar{t}_2 \rk = T \lk \bar{t}_3 \rk - T \lk \bar{t}_1 \rk =
\frac{1}{2} - \lk -\frac{1}{2} \rk = 1.
\ee

Let us now investigate how this change of the twist is distributed over the
various singular points at $t = \bar{t}_2$.
\medskip

It is easy to see that the intersection point of $C_1$ and $C_5$ at $\lk x,y,z
\rk = \lk 0,0,0 \rk$ and $t=\bar{t}_2$ does not change the twist. This is because
the line segments $C_1$ and $C_5$ do not contribute to the twist at both $t =
\bar{t}_1 < \bar{t}_2$ and $t = \bar{t}_3 > \bar{t}_2$ since the framing vector
$\hat{n} (s)$ does not change along these line segments (see also the explicit
calculations in Appendix B).
The change of twist at
$t = \bar{t}_2$ entirely arises from the twisting points at $\lk x,y,z \rk = \lk
\pm a,0,0 \rk $ due to the reflection of the line segments $C_4$ and $C_6$ at
the $x-z$-plane. 
Indeed, we could carry out the
change of the vortex loop at $t = \bar{t}_2$ in two steps: First we could deform
the line segment $C_5$ such that it intersects the line segment $C_1$ keeping the
boundary points of $C_5$ fixed (see fig. \ref{fig9} (a)), so that the line segments $C_4$
and $C_6$ remain unchanged. This intersection of $C_5$ and $C_1$ would still
change the writhing number by $\Delta W \lk t_2 \rk ^{(i)} = -2$ but at the same
time leave the torsion unchanged (since $C_5$ and $C_1$ to not contribute to $T$).
Second, to produce the vortex configuration at $t = \bar{t}_2 + \varepsilon , \varepsilon  >
0$ we deform the line segment $C_5$ to its final form by reflecting the end
points of $C_5$ at the $x-z$-plane, which will also reflect the line segments $C_4$
and $C_6$ (including their framings) see figs. \ref{fig9} (b), (c). This second change of the
vortex loop does not only change the writhing number, namely by $\Delta W \lk
\bar{t}_2 \rk ^t (a) + \Delta W \lk \bar{t}_2 \rk ^t (-a) = \frac{1}{2} +
\frac{1}{2} = 1$ as discussed above, but also the twist. 
\medskip

One easily convinces oneself that in the second step only the sign of the
twists along $C_4$ and $C_6$ changes. This can be seen as follows. The
reflection at the $x-z$-plane reverses only the sign of the $y$-component of a
vector (while the $x$-and $z$-components are unchanged). The vectors $\hat n (s), 
\dot{\hat n}(s)$ of the framing along $C_4$ and $C_6$ are parallel to $x-z$-plane and
are consequently not changed by the reflection at the $x-z$-plane. Thus also
$\hat n (s) \times  \dot{\hat n}(s)$ (which is an axial vector) does not change
although it is parallel to the $y$-axis. What changes in the second step is the
sign of the $y$-component of the vector $\vec{r}(s)$. Thus the integrand in the
twist (\ref{G89}), and hence the twists along $C_4$ and $C_6$ themselves, change
the sign:

\be
T (C_4 ((\bar t _2 -\varepsilon)) = - \frac{1}{4} \Rightarrow T (C_4 (\bar t
_2 + \varepsilon )) = + \frac{1}{4},\nonumber\\ 
T (C_6 (\bar t_2 -\varepsilon )) = -
\frac{1}{4} \Rightarrow T (C_6 (\bar t _2 + \varepsilon)) = + \frac{1}{4} . 
\ee

Consequently each of the twisting points at $\lk x,y,z \rk = \lk 0,0,0 \rk $
change the twist by $\Delta T = + \frac{1}{2}$ , so that the total change of
the twist at $t = \bar{t}_2$ is $\Delta T = 1$.
This property justifies the name ``twisting point''. Twisting points change not
only the writhing number by $|\Delta W | < 2$ but, depending on the framing, may also 
change the twist, while
intersection points change the writhing number by $| \Delta W | = 2$
without changing the twist. 
In view of eq.~(\ref{G88}) the intersection points change also the self-linking
number by $\Delta S L = \Delta W = \pm 2$. 
\medskip

In fig. \ref{fig7} (a) and (b) the framings of the vortex loops at $t= \bar{t}_1$ and $t =
\bar{t}_3$ have been chosen in such a way that the framing of the line-segment
$C_5$ does not change as $C_5$ moves from $y=-a$ to $y=a$ thereby intersecting
$C_1$ at $y=0$. As discussed above, with these framings during the evolution
from $t=\bar{t}_1$ to $t=\bar{t}_3$ the writhing number changes from $W \lk \bar{t}_1 \rk = \frac{1}{2}$ to
$W\lk \bar{t}_3 \rk = - \frac{1}{2}$ and the twist changes from $T\lk \bar{t}_1
\rk =  - \frac{1}{2}$ to $T \lk \bar{t}_3 \rk = \frac{1}{2}$ so that the self-linking number
remains unchanged.
The ribbons of the framings at both $t= \bar{t}_1 , \bar{t}_3$ shown in fig.
\ref{fig7} (a), (b) have vanishing self-linking numbers. 
\medskip

To illustrate the frame-independence 
of the writhing number and the frame-dependence of the twist and of the
self-linking number let us choose the alternative framing of the vortex loop at
$t = \bar{t}_1$ shown in fig. \ref{fig7} (c). With this framing the twist of the vortex
loop at $t=\bar{t}_1$ is $T  ( \bar t _1 )= \frac{1}{2}$ while the writhing
number is still $W (\bar t _1 )=
\frac{1}{2}$, as it should since $W$ is framing-independent. The ribbon of the
framing at $t = \bar{t}_1$ (shown in fig. \ref{fig7} (c)) is now
a twisted band with the self-linking number $ S L
(\bar t _1 ) =1$.
We can now turn the
framed vortex configuration at $t=\bar{t}_1$ shown in fig. \ref{fig7} (c) into the one at $ t =
\bar{t}_3 $ shown in fig. \ref{fig7} (b) by rotating the yoke $(C_4 , C_5 , C_6)$ (including the
corresponding framing) around the x-axis by angle $\pi$. During this
process the twist does not change i.e. $T\lk \bar{t}_1 \rk = T \lk \bar{t}_3 \rk
= \frac{1}{2}$ but the writhing number changes as in the previous case from $W \lk
\bar{t}_1 \rk = \frac{1}{2}$ to $W\lk \bar{t}_3 \rk = - \frac{1}{2}$.
Accordingly the self-linking number changes from $S L \lk \bar{t}_1 \rk = 1$ to
$S L \lk \bar{t}_3 \rk = 0$. 

In sect. 4 I have shown that only discontinuous changes in the writhing number
(during the time evolution) of the vortex loop contribute to the topological
charge. (Continuous changes are fully taken into account by the initial and final
value of the writhing number, i.e. by $\lk W \lk t_f \rk - W \lk t_i
\rk\rk $, see eq.~(\ref{G86}).)When the yoke $\lk C_4 , C_5 , C_6 \rk$ is rotated by an angle $\pi$ around the
$x$-axis the writhing number changes continuously except at the time when $C_5$
crosses (the end of) $C_1$. This represents a crossing of the ``full'' line
$C_5$ with the ``half-line'' $C_1$ (see fig. \ref{fig3} (b)) which changes the writhing
number by $\Delta W = -1$. Furthermore when $C_5$ crosses $C_1$ the vortex loop
is planar so that its writhing number vanishes. Thus when $C_5$ intersects
$C_1$ the writhing number changes from $W= \frac{1}{2}$ to $W=-\frac{1}{2}$ in
accord with our previous findings. Note that in the present case the total
change of the writhing number comes from the ``half-intersection'' point $\lk 0,
0, a \rk$ of
$C_5$ and $C_1$ which, in fact, is a twisting point. There
are no further twisting points in this case since there are no further singular
changes of the vortex loop. Note also when $C_5$ passes $C_1$ (keeping the
framing fixed) the twist does not change, $\Delta T = 0$, but the self-linking
number changes by $\Delta SL \lk C \rk = 1$, in agreement with our
previous findings.

\no
\subsection{Non-oriented center vortices}
\bi

As discussed in sect. 3 center vortex sheets with non-zero topological charge
are necessarily non-oriented and carry magnetic monopole loops at the interface
between vortex patches of different orientations. One can convince
oneself that the above given analysis of the topological charge of generic
center vortices in terms of the writhing number remains valid for non-oriented
vortices, i.e. in the presence of magnetic monopoles. On a (non-oriented)
generic vortex sheet the monopole loops evolve in time (like the vortex sheet)
and show up at a fixed time on the vortex loop as monopole-anti-monopole pairs.
Both monopole and anti-monopole change the direction of the flux of the vortex
loops. The writhing number depends not only on the shape but also on the
orientation of the loops and fully accounts also for the topology of generic
non-oriented center vortices as we will now demonstrate explicitly. 
\medskip

For this purpose we convert the oriented center vortex shown in fig. \ref{fig4} into a
non-oriented one by putting a magnetic monopole loop in the $z-t$-plane.  
The time-evolution of this non-oriented vortex in the continuum is illustrated
in fig.\ref{fig10}, which is the counter part of fig.\ref{fig6}. 
For calculational simplicity we have chosen the monopole loop as follows:
At time $t'_1$ with $\bar t _1 < t'_1 < \bar t _2$ a magnetic monopole-antimonopole
pair is created at $z=0$. Monopole and antimonopole then run away from each
other until they reach the positions $z=a$ and $z=-a$, respectively, at a time
$t_1 ''$ with $t'_1 < t_1 '' < \bar t _2$. Then monopole and anti-monopole
keep their position in space up to a time $t_2 ''$ with $\bar t _2 < t_2 '' <
\bar{t}_3$ and subsequently approach each other until they annihilate at a time
$t'_2$ with $t_2 ''<t_2 '<\bar t _3$.
\medskip

The presence of the monopole-antimonopole pair on the vortex loop does not give
rise to additional singular changes of the writhing number as function of the
time. However, the magnetic monopoles have reversed the orientation of (the
relevant part of) the line
segment $C_1$ (see figs. \ref{fig11}, \ref{fig10}) for times $t \in \lke t_1 '' , t_2 '' \rke$ and
thus in particular at time $t=\bar t _2$ when $C_1$ crosses $C_5$. We therefore
expect that the contribution of the intersection (crossing) point at $t = \bar
t _2, \lk x,y,z \rk = \lk 0,0,0 \rk$ to the change of the writhing number
$\Delta W (t_2 )^i (0)$ (see eq.~(\ref{G97})) has opposite sign for the non-oriented
vortex. Indeed, the contribution of the intersection point is given by (see
Appendix A, eq.~(\ref{G123})): 

\be
\Delta W (\bar t _2 ) ^i (0) = 4 \lim \limits_{b \rightarrow 0} L \lk C_1 , C_5
\rk ,
\ee

where $L \lk C_1 , C_5 \rk$ is the Gaussian linking integral for open
paths defined by eq.~(\ref{G110})
in Appendix A and 
$b$ denotes the distance of the line-segment $C_5$ from the $x$-axis. Changing the orientation of $C_1$ changes the sign of $L \lk C_1 ,C_5 \rk$ and in
view of eq.~(\ref{G97}) we obtain

\be
\Delta W (\bar{t}_2 )^i (0) = 2.
\ee

The changes of the writhing number due to the two twisting points at $t=\bar t
_2 , \lk x,y,z \rk = \lk \pm a,0,0 \rk , \Delta W \lk \bar t _2 \rk ^t (\pm a
)$, do not receive contributions from the line segment $C_1$ and are hence not
changed by the addition of the magnetic monopoles on $C_1$. Hence we find for the change
of the writhing number at $t = \bar t _2$

\be
\begin{array}{lllll}
\Delta W (\bar t _2 ) & = \Delta W (\bar t _2 ) ^i (0) & + \Delta W (\bar t _2 )
^t (a) & + \Delta W (\bar t _2 ) ^t (-a)\nonumber\\
& = 2 & + \frac{1}{2} & + \frac{1}{2}  & = 3.
\end{array}
\ee

The writhing numbers of the vortex loop at $t = \bar t _1 , \bar t _3$
are not changed by the addition of the monopole loop since the latter does not
exist at these times. Thus for the non-oriented center vortex shown in fig.
\ref{fig10}
the total change of the writhing number during the time evolution of the vortex
is given by

\be
\begin{array}{lllll}
\Delta W & = W (\bar t _1 ) & + \Delta W (\bar t _2 ) & - W (\bar t _ 3)\nonumber\\
& = \frac{1}{2} & +3 & + \frac{1}{2} & = 4
\end{array}
\ee

which, in view of eq.~(\ref{G86}), corresponds to a Pontryagin index $\nu = 1$.
Thus the non-oriented center vortex shown in fig. \ref{fig10} is topologically
equivalent to an instanton. 
\medskip

A final remark is in order: In the above
investigation of non-oriented center vortices we have tacitly extended the
writhing number to non-oriented loops. In fact, since the writhing number, as well
as the twist and the Gaussian self-linking number, are defined by loop
integrals,
these quantities can be straightforwardly generalised to non-oriented loops.
One should, however, be aware that for non-(globally) oriented loops the
self-linking number (\ref{G87}), (\ref{G82}) is not necessarily a topological invariant and the
relation (\ref{G88}) need not to hold.   

\no
\section{Summary and conclusions}
\bi

In the present paper we have studied the topological properties of center
vortices. The topological charge of center vortices is given by (a quarter of)
their self-intersection number, which, however, vanishes for closed oriented
vortex sheets . Topologically non-trivial vortex sheets are not globally oriented but consist of oriented vortex patches,
which are joined by magnetic monopole loops. These monopole loops are irrelevant
for the confining properties of center vortices as measured by their
contributions to the Wilson loop but absolutely crucial for their topological
properties.
\medskip

The intersection number of vortex sheets receives contributions from two types
of singular points: transversal intersection points, contributing $\pm
\frac{1}{2}$ to the topological charge, and twisting points which give
contributions of modulus smaller than $\frac{1}{2}$. 
\medskip

In ordinary 3-space, generic center vortices represent (in general
time-dependent) closed magnetic flux loops. I have shown that the topological
charge of these center vortices can be entirely expressed by the change of the
writhing number of the vortex loops during the time-evolution. I have also
demonstrated that transversal intersection points of the 2-dimensional vortex
sheet in D=4 correspond to a crossing of two line segments of the vortex loop in
D=3. Furthermore the twisting points of the vortex sheets in D=4 show up as
rotations and, in particular, reflections of line segments of the vortex loop in
D=3. I believe
that the present analysis of the topological properties of center vortices in
terms of the writhing number of time-dependent flux loops in D=3 is more
transparent, although not necessarily calculationally simpler, than the analysis
of the self-intersection number of closed vortex sheets in D=4. 
\bigskip

\no
{\Large \bf Acknowledgements:}
\bi

\bigskip

\begin{figure}
\centerline{
\epsfysize=8.5cm
\epsffile{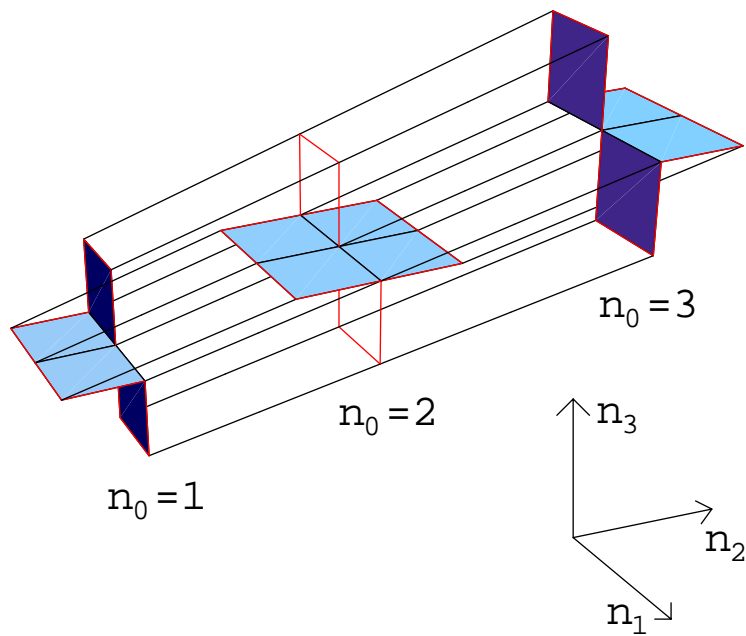}
}
\caption{Sample vortex surface configuration taken from \cite{[Eng2]}. At each lattice time
$t=n_0 a$ (a-lattice spacing), shaded plaquettes are part of the vortex surface. These
plaquettes are furthermore connected to plaquettes running in time
direction; their location can be inferred most easily by keeping in
mind that each link of the configuration is connected to exactly
two plaquettes (i.e. the surface is closed and contains no intersection
lines). Note that the two non-shaded plaquettes at $n_0 =2$ are {\em not}
part of the vortex; only the two sets of three links bounding them are.
These are slices at $n_0 =2$ of surface segments running in time
direction from $n_0 =1$ through to $n_0 =3$. Sliced at $n_0 =2$, these
surface segments show up as lines. Furthermore, by successively assigning
orientations to all plaquettes, one can convince oneself that the
configuration is orientable. The vortex image was generated by means of a
MATHEMATICA routine provided by R.~Bertle
and M.~Faber.}
\label{fig4}
\end{figure}

\begin{figure}
\centerline{
\epsfysize=12.0cm
\epsffile{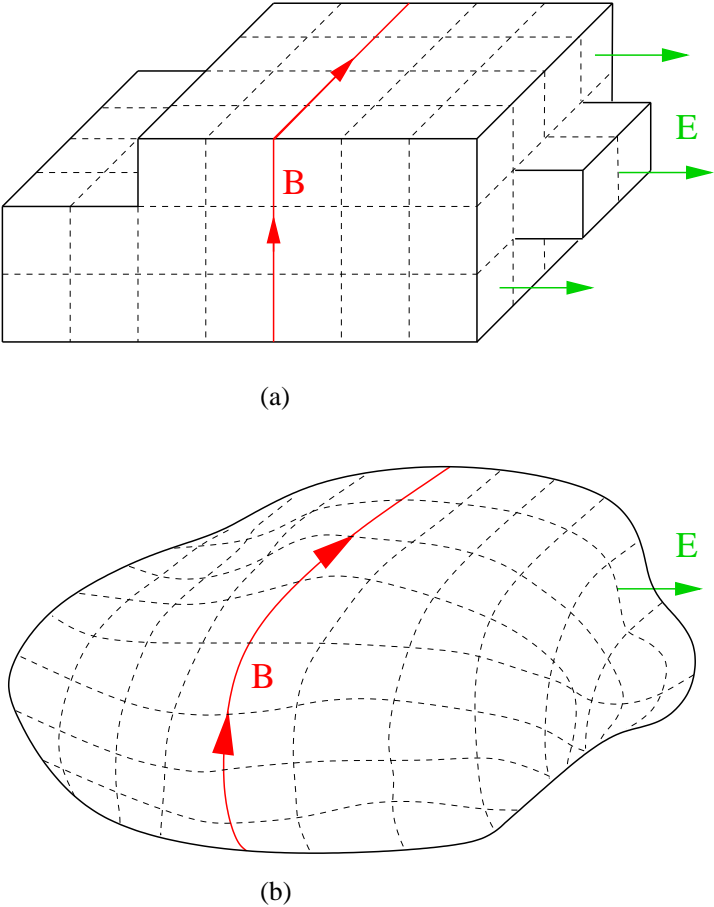}
}
\caption{(a) Lattice realization of the smooth continuum center vortex
  configuration shown in (b). The use of discrete time intervals on the
  lattice gives rise to purely spatial vortex surface patches, which should
  be considered as lattice artifacts.}
\label{fig5}
\end{figure}

The author is grateful to M. Engelhardt for useful discussions, in particular,
on the center vortex configuration shown in fig.\ref{fig4}., and to J. Gattnar
for preparing the tex files of the other figures. He also acknowledges discussions
with T. Tok. 
\bigskip

{\Large \bf Appendix A: Calculation of the writhing number}
\bi

\begin{figure}
\centerline{
\epsfysize=4.0cm
\epsffile{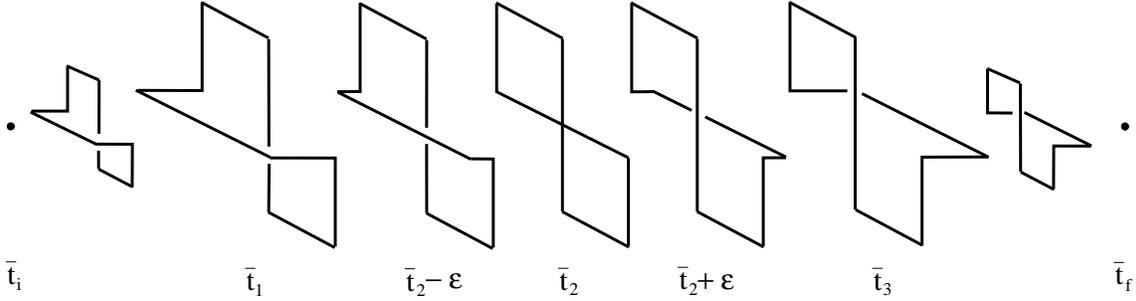}
}
  \caption{Snap shots at characteristic time instants of the continuum center 
    vortex loop whose lattice realization is shown in 
     \ref{fig4}.}
    \label{fig6}
\end{figure}

\begin{figure}
\centerline{
\epsfysize=8.5cm
\epsffile{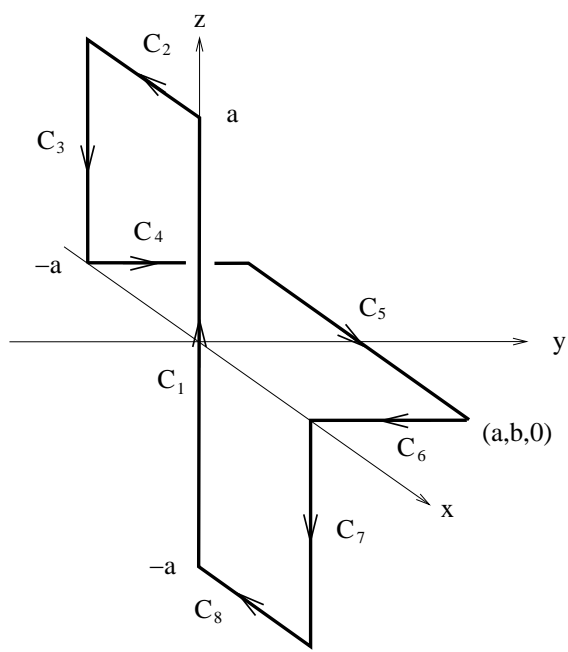}
}
\caption{Definition of the line-segments $C_i$ used in the calculation of the
writhing number and twist of the vortex loop at $t = \bar t _3$ shown in fig. \ref{fig6}.}
\label{fig11}
\end{figure}

In the following we calculate the changes in the writhing number of the vortex
configuration shown in fig. \ref{fig6}. We begin by calculating the writhing number
of the final field configuration at $t = \bar{t}_3$. For later convenience we
will, however, start by calculating the writhing number of the slightly
generalised loop configuration C shown in fig. \ref{fig11}.  This loop C has a
cubistic form. It consist of straight line segments $C_i , i=, ...,8,$ which
are prallel to one of the Cartesian coordinate axis and are a distance $a$ apart
form the latter, except for the line segment $C_1$, which sits on the $z$-asis,
and the line segment $C_5$, which has a distant $b$ form the $x$-axis. For $b=a$  
the loop C reduces to the
vortex configuration at $t = \bar{t}_3$ shown in fig. \ref{fig6}.
\medskip

It is convenient to break the loop C into the straight line
segments $C_i$, i.e. $C=\bigcup \limits_{i=1}^8 C_i$. The writhing number is then given by 

\be
\label{G109}W \lk C (b) \rk = L (C,C) = \sum \limits_{i,j=1} ^8  L \lk C_{i,} C_j \rk
\ee

where

\be
\label{G110}L \lk C_{i,}C_j \rk = \frac {1}{4\pi} \int \limits_{C_i} ( d \vec{x} \times \int
\limits_{C_j} d\vec{x'} ) \cdot \frac {\lk \vec{x} - \vec{x'} \rk }{\left |
\vec{x} - \vec{x'} \right | ^3}
\ee

is the Gaussian linking integral for open paths $C_i$. Due to the cross product
the integrand obviously vanishes when the two-line segments are parallel, i.e.
$d \vec{x} \left |\right | d\vec{x'}$, so that $L \lk C_{i}, C_i \rk = 0$. Since
furthermore $L \lk C_{i}, C_j \rk = L \lk C_{j}, C_i \rk$, we obtain 

\be
\label{G111}W (C) = 2\sum \limits_{i<j} L \lk C_{i}, C_j \rk.
\ee

By the same argument pairs of  parallel paths do not contribute, for example $L
\lk C_{2},C_5 \rk =0$. Furthermore, the integrand, containing the product $\lk d
\vec{x} \times d\vec{x'} \rk \cdot \lk \vec{x} - \vec{x'} \rk $, vanishes when
the three vectors $d \vec{x} , d \vec{x'} , \lk \vec{x} - \vec{x'} \rk $ are in
the same plane. This happens for any two pairs of paths within the following sets of
paths $\lk C_{1},C_{2},C_{3},C_{7},C_{8}\rk ,\lk C_{4},C_{5},C_{6} \rk
\lk C_{3},C_4\rk ,\lk C_{6},C_{7}\rk $. Thus the only non-vanishing
contributions to the writhing number are given by

\be
\label{G112}W(C) & = & 2 \big[ L \lk C_ {1},C_4 \rk + L \lk C_{1},C_5 \rk \nonumber\\
& + & L \lk C_{1},C_6 \rk + L \lk C_{2},C_4 \rk + L \lk C_{2},C_6 \rk
\nonumber\\
& + & L \lk C_{3},C_5 \rk + L \lk C_{3},C_6 \rk + L \lk C_{4},C_7 \rk
\nonumber\\
&+ & L \lk C_{4},C_8 \rk + L \lk C_{5},C_7 \rk + L \lk C_{6},C_8 \rk \big].
\ee

\medskip

Since the paths $C_i$ are all parallel to one of the coordinate axes, we use the
corresponding Cartesian-coordinates to parametrise these paths 

\be
\label{G113}
\begin{array}{ll} 
C_1 :\vec{r}  =  z \vec{e}_z , & \int \limits_{C_1} d\vec{r} =   
\vec{e}_z \int \limits_{-a} ^a dz,\\  
C_2 :\vec{r} =  x \vec{e}_x + a \vec{e}_z, & \int \limits_{C_2}  d \vec{r}
= \vec{e}_x   \int \limits_0 ^{-a}  dx,\\
C_3 :\vec{r} = -a\vec{e}_x + z \vec{e}_z, & \int \limits_{C_3} d \vec{r} =
\vec{e}_z \int \limits_{y} ^0 dz,\\
C_4 :\vec{r} =  -a \vec{e}_x + y \vec{e}_y, & \int \limits_{C_4} d \vec{r} =
\vec{e}_y \int \limits_0 ^b d y,\\
C_5 :\vec{r} =  x \vec{e}_x + b \vec{e}_y, & \int \limits_{C_5} d \vec{r} =
\vec{e}_x \int \limits_{-a} ^a dx,\\ 
C_6 :\vec{r} =  a \vec{e}_x + y \vec{e}_y, & \int \limits_{C_6} d \vec{r} =
\vec{e}_y \int \limits_b ^0 dy,\\ 
C_7 :\vec{r} = a \vec{e}_x + z \vec{e}_z, & \int \limits_{C_7} d \vec{r} =
\vec{e}_z \int \limits_0 ^{-a} dz,\\ 
C_8 :\vec{r} =  x \vec{e}_x -a \vec{e}_z, & \int \limits_{C_8} d \vec{r} =
\vec{e}_x  \int \limits_a ^0 dx. 
\end{array}
\ee

It is then straightforward to calculate the linking integrals $L \lk C_{i,}
C_j \rk$. For notational simplicity let us introduce the following abbreviations

\be
\label{G114}L_{ij} & = & 4\pi L \lk C_{i,} C_j \rk ,\nonumber\\ 
f (x,y,z) & = &  \lk x^2 + y^2 + z^2 \rk ^{-3/2}
\ee

Rescaling all coordinates by $a$, i.e. $x/a \rightarrow x \hspace{0.2cm}$ etc.,
we find the explicit expressions

\be
\label{G115}
L_{14}= - \int\limits_0 ^{\frac{b}{a}} dy \int\limits_0 ^1 dz f \lk 1,y,z
\rk, \nonumber
& L_{15}=  - \frac{b}{a} \int\limits_{-1} ^1 dx \int\limits_{-1} ^1 dz f \lk
x, \nonumber
\frac{b}{a}, z \rk\\\nonumber
L_{16}= \int \limits_{b/a} ^0 dy \int \limits_{-1} ^1 dz f \lk 1,y,z \rk, \nonumber
& L_{24}= \int \limits_0^{-1} dx \int \limits_0 ^{b/a} dy f \lk x+1,y,1 \rk,
\nonumber 
\\
L_{26}=\int \limits_0 ^1 dx \int \limits_0 ^{b/a} dy f \lk x+1,y,1 \rk, \nonumber
& L_{35}= \frac{b}{a} \int \limits_0 ^2 dx \int \limits_0 ^1 dz f \lk x,
\nonumber
\frac{b}{a} ,z \rk, 
\\\nonumber
L_{36}= 2 \int \limits_0 ^{b/a} dy \int \limits_0 ^1 dz f \lk 2,y,z \rk ,\nonumber
& L_{47}= -2 \int \limits_0 ^{b/a} dy \int \limits_0 ^{-1} dz f \lk 2,y,z \rk
,
\nonumber
\\
L_{48}= -\int \limits_1 ^0 dx \int \limits_0 ^{b/a} dy f \lk x+1,y,1 \rk ,\nonumber
& L_{57}=\frac{b}{a}  \int \limits_{-2} ^0 dx \int \limits_0 ^1 dz f \lk
x,\nonumber
\frac{b}{a},z \rk, \\
L_{68} = - \int \limits_{-1} ^0 dx \int \limits_0 ^{b/a} dy f \lk x,y,1 \rk.
\ee

Let us first consider the writhing number of the final field configuration
at $t = \bar{t}_3$ (see fig. \ref{fig6}) for which $b=a$. In this case all linking integrals can be
reduced to the three following integrals 

\be
\label{G116}I = \int \limits_0 ^1 dx \int \limits_0 ^1 dy f \lk x,y,1 \rk , \nonumber\\
K = \int \limits_0 ^1 dx \int \limits_1 ^2 dy f \lk x,y,1 \rk , \nonumber\\
M = \int \limits_0 ^1 dx \int \limits_0 ^1 dy f \lk x,y,2 \rk .
\ee

One finds

\be
\label{G117}
\begin{array}{lllll}
L_{14} = -2I, & L_{15} = -4I,& L_{16} = -2I,\nonumber\\
L_{24} = -I, & L_{26}  =  K, & L_{35} =  I + K, & L_{36} = 2M,\nonumber\\
L_{47} = 2M, &  L_{48}  =  K, & L_{57} =  I + K,&  L_{68} = -I.
\end{array}
\ee

With this result the writhing number becomes $ \lk C \equiv C \lk b = a  \rk
\rk$  

\be
\label{G118}W (C) = \frac{8}{4\pi} \lke -2I+K+M \rke.
\ee

The integrals $I,K,M$ can be trivially taken by noticing that 

\be
\label{G119}f \lk x,y,z \rk = \lk x^2 + y^2 + z^2 \rk ^{-3/2} = \frac{1}{y^2 + z^2}
\frac{d}{dx} \frac{x}{\sqrt{x^2 + y^2 + z^2}}.
\ee

One finds

\be
I = \frac{\pi}{6},\hspace{0.5cm} K = \arctan \lk \frac{1}{3} \sqrt 6 \rk -
\frac{\pi}{6}, \hspace{0.5cm}M  = \frac{1}{2} \arctan \lk \frac{1}{12} \sqrt 6
\rk 
\ee

so that the writhing number becomes 

\be
\label{G120}W(C)= -1 + \frac{2}{\pi} \lke \arctan \lk \frac{1}{3} \sqrt{6} \rk + \frac{1}{2}\arctan \lk
\frac{1}{12} \sqrt{6}\rk \rke.
\ee

Using the addition theorem for the arcus-tangent function one eventually finds 

\be
\label{G121} W \lk \bar t _3 \rk = W \lk C \rk = - \frac{1}{2}.
\ee

The calculation of the writhing number of the vortex configuration at $t = \bar
t _1$ proceeds completely analogously. The writhing number of the vortex loops
at $t = \bar t _1$ is still given by eq.~(\ref{G112}). Furthermore the
line segments $C_1 , C_2 , C_3 , C_7 $ and $C_8$ are the same for the vortex
loops at $t = \bar t _3$ and $t = \bar t _1$. What is different are the line
segments $C_4 , C_5$ and $C_6$. The changes of these line segments can be seen
to merely change the signs of {\it{all}} contributing linking integrals $L_{ij}$
quoted in eq.~(\ref{G115}), so that we obtain 

\be
W (\bar t _1 ) = - W (\bar t _3 ) = \frac{1}{2}.
\ee

Let us now investigate how the writhing numbers changes at $ t = \bar{t}_2$ as $C_5$
crosses $C_1$. There is obviously an intersection point at $\lk
x,y,z \rk = \lk 0,0,0 \rk$ at which $W(C)$ changes by $\left | \Delta W \right |
= 2$. With the above notation the change in the writhing number at
$ t= \bar{t}_2$  when $C_5$ crosses $C_1$ is given by

\be
\label{G122}\Delta W \lk \bar{t}_2 \rk = \lim \limits_{b \rightarrow 0} \lke W
\lk C \lk b \rk \rk 
-W \lk C \lk -b \rk \rk \rke . 
\ee

\medskip

Obviously only such linking integrals $L_{ij}$ contribute to $\Delta W \lk
\bar{t}_2 \rk$ (\ref{G122})  which are odd in $b$ and non-vanishing
in the limit $b \rightarrow 0$. Since the paths $C_4$ and $C_6$ vanish in the
limit $b \rightarrow 0$ all linking integrals containing $C_4$ or $C_6$ vanish
$L \lk C_4 , C_i \rk = L \lk C_6 , C_i \rk = 0$ for all $i$. The remaining non-zero
linking integrals are in fact odd in $b$ and we obtain from eq.~(\ref{G112}) for the
change in the writhing number 

\be
\label{G123}\Delta W \lk \bar{t}_2 \rk  & = &  4 \lim \limits_{b \rightarrow 0 } \lke L \lk C_1 ,
C_5 \rk + L \lk C_3 , C_5  \rk + L \lk C_5 , C_7 \rk \rke \nonumber\\
& = & \Delta W \lk \bar{t}_2 \rk ^i (0) + \Delta W \lk \bar{t}_i \rk ^t (-a) +
\Delta W \lk \bar{t}_i \rk ^t (a)
\ee

Here the first, second and third terms represent the contributions to the change in
the writhing number from the intersection point at $\lk 0,0,0 \rk$ and the twisting
points at $ \lk -a,0,0 \rk $ and $\lk a,0,0 \rk$, respectively. Using the
explicit expressions for the linking integrals given in eq.~(\ref{G115}) and taking
the limit $b \rightarrow 0$ one finds 

\be
\label{G124}\Delta W \lk \bar{t}_2 \rk ^i (0) & = & -2,\nonumber\\
\Delta W \lk \bar{t}_2 \rk ^t (\mp a) & = & \frac {1}{2}. 
\ee 

The intersection point changes the writhing number by $(-2)$ while the twisting
points change it by $\frac{1}{2}$. The total change of the writhing number at $ t=
\bar{t}_2 $ is then given by 

\be
\label{G125}\Delta W \lk \bar{t}_2 \rk = -1.
\ee 

{\Large \bf Appendix B: The twist}
\bigskip

As a cross check we calculate in the following the twist of the final vortex
configuration shown in fig. \ref{fig6} at $t= \bar{t}_3 $, see also fig.
\ref{fig11}. Note that due
to eq.~(\ref{G88}) the sum of the twist
and the writhing number has to be integer-valued.
\medskip

\begin{figure}
\centerline{
\epsfysize=20.0cm
\epsffile{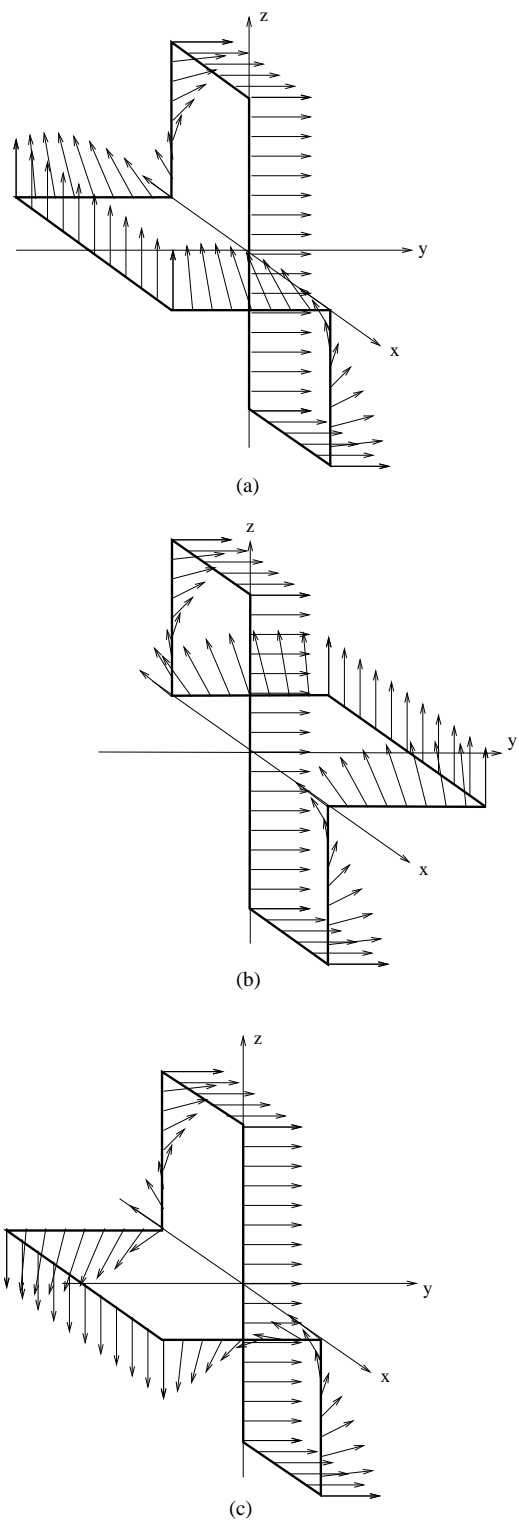}
}
\caption{Framings of the vortex loop shown in fig. \ref{fig6}; (a) and (b) show framings of
the vortex loop at $t= \bar{t}_1$ and $t=\bar{t}_3$, respectively, with
vanishing self-linking number, $SL(C)=0$.  (c) framing of the vortex loop at $t=\bar{t}_1$
with $SL(C)=1$.}
\label{fig7}
\end{figure}

\begin{figure}
\centerline{
\epsfysize=8.0cm
\epsffile{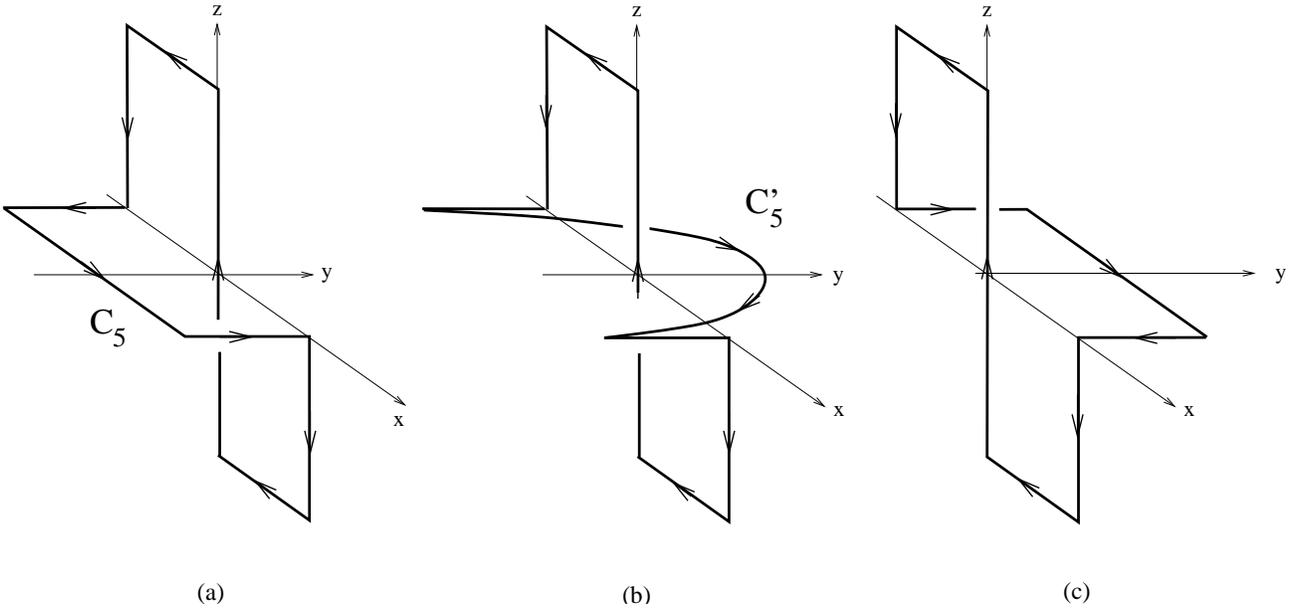}
}
\caption{(a) The vortex loop of Fig. \ref{fig6} at time $t=\bar{t}_1$ (b) shows
the same loop as in (a) except that the line-segment $C_5$ has been deformed to
the line-segments $C'_5$ thereby intersecting the line-segment $C_1$. (c) The
vortex loop of Fig. \ref{fig6} at time $t=\bar{t}_3$.}
\label{fig9}
\end{figure}

\begin{figure}
\centerline{
\epsfysize=5.0cm
\epsffile{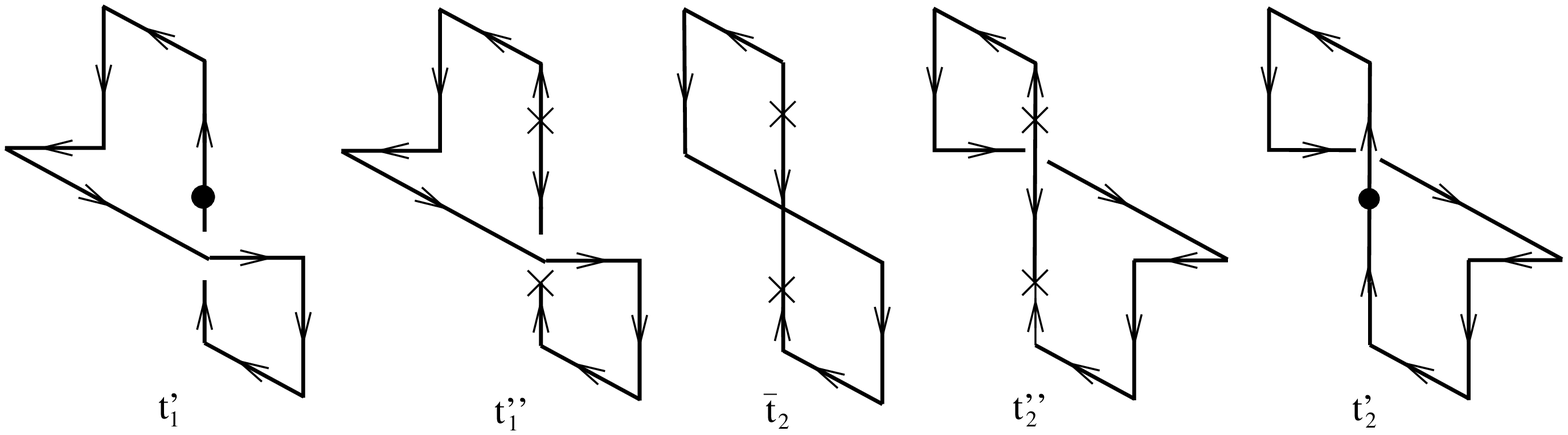}
}
\caption{Snap shots at characteristic time instants of the continuum center vortex
loop whose lattice realization is obtained from the center vortex
    configuration shown in fig. \ref{fig4} if a magnetic monopole loop is added
    in the $n_0 -n_3$-plane around the point $\lk n_0 , n_1 , n_2 , n_3 \rk =
    \lk 2,0,0,0\rk$. The crosses denote the positions of the monopole and
    anti-monopole. The fat dots indicate the creation and annihilation of a
    monopole-anti-monopole pair.}
\label{fig10}
\end{figure}

To evaluate the twist we use the same parametrisation of
the path as in the evaluation of the writhing number, see Appendix A. Furthermore
we use the framing shown in fig. \ref{fig7}. This framing of the path, 
i.e. $\hat{n} (s)$, does not change along
the loop segments $C_1 , C_2 , C_5 , C_8,$ which hence do not contribute to the
twist. At the remaining loop segments the framing is chosen as

\be
\label{G126}\begin{array}{ll}
C_3  : & \hat{n}  (z) = \vec{e}_y \cos \pi \frac{a-z}{za} - \vec{e}_x \sin \pi
\frac{a-z}{2a},\\
C_4 : & \hat{n} (y) = - \vec{e}_x \cos \frac{\pi y}{2a} + \vec{e}_z \sin
\frac{\pi y}{2a},\\
C_6 : & \hat{n}  (y) = \vec{e}_z \sin \frac{\pi y}{2a} - \vec{e}_x \cos \frac{\pi
y}{2a},\\
C_7 : & \hat{n} (z) = - \vec{e}_x \cos \frac{\pi z}{za} - \vec{e}_y \sin \frac{\pi
z}{2a}.
\end{array}
\ee

Straightforward evaluation of the twist integrals yields the following results

\be
\begin{array}{lllll}
\label{G127}T \lk C_3 ,\hat{n} \rk & = &  - \frac{1}{4}, T \lk C_4
,\hat{n} \rk & = & \frac{1}{4},\\
T \lk C_6 , \hat{n} \rk & = & \frac{1}{4}, T \lk C_7 ,\hat n \rk & = &
\frac{1}{4}.
\end{array}
\ee

The total twist of the vortex loop at $ t= \bar{t}_3$ is then given by

\be
\begin{array}{llllll}
\label{G128}T \lk C \lk \bar{t}_3 \rk \rk & =  T \lk C_3 ,\hat{n}\rk & + T
\lk C_4 , \hat{n} \rk & + T \lk C_6 ,\hat{n}\rk & + \lk C_7 , \hat{n} \rk \\
& = - \frac{1}{4} & + \frac{1}{4} & + \frac{1}{4} & + \frac{1}{4} & =
\frac{1}{2}.
\end{array}
\ee

\newpage

\begin{thebibliography}{99}

\bibitem{[Hooft]}G.`t Hooft, Nucl. Phys. {\bf{B138}} (1978) 1;\\
Y. Aharonov, A. Casher and S. Yankielowicz, Nucl. Phys. {\bf{B146}}
(1978) 256;\\ J. M. Cornwall, Nucl. Phys. {\bf{B157}} (1979) 392\\
G. Mack and V. B. Petkova, Ann. Phys. (NY) {\bf{123}} (1979) 442;\\
G. Mack, Phys. Rev. Lett. {\bf{45}} (1980) 1378;\\
G. Mack and V. B. Petkova, Ann. Phys. (NY) {\bf{125}} (1989) 117;\\
G. Mack, in: {\it{Recent Developments in Gauge Theories}} , eds. G. 't Hooft et
al. (Plenum, New York, 1980); \\ G. Mack and E. Pietarinen, Nucl. Phys.
{\bf{B205}} [FS5] (1982) 141\\
H. B. Nielsen and P. Olesen, Nucl. Phys. {\bf{B160}} (1979) 380;\\ H.
Ambj\o rn and P. Olesen, Nucl. Phys. {\bf{B170}} [FS1] (1980) 60;\\ J.
J. Ambj\o rn and P. Olesen, Nucl. Phys. {\bf{B170}} [FS1] (1980) 265;\\
E. T. Tomboulis, Phys. Rev. {\bf{D 23}} (1981) 2371

\bibitem{[Del2]}Del Debbio, M. Faber, J. Greensite, $\breve S$. Olejnik, Phys.
Rev. {\bf{D55}} (1997) 2298

\bibitem{[Lang]}K. Langfeld, H. Reinhardt, O. Tennert, Phys. Lett.
{\bf{B419}} (1998) 317

\bibitem{[Ldel]}L. Del Debbio. M. Faber, J. Giedt, J. Greensite and $\breve S$ .
Olejnik, Phys. Rev. {\bf{D 58}} (1998) 094501

\bibitem{[Klang]}K. Langfeld, O. Tennert, M. Engelhardt and H. Reinhardt, Phys.
Lett. {\bf{B542}} (1999) 301, M. Engelhardt, K. Langfeld, H. Reinhardt
and O. Tennert, Phys. Rev. {\bf{D61}} (2000) 054504 

\bibitem{[Kova]}T. G. Kovacs, E. T. Tomboulis, Phys. Rev. Lett.
{\bf{85}} (2000) 704

\bibitem{[Forc]}P. de Forcrand and M. D`Elia, Phys. Rev. Lett. {\bf{82}}
(1999) 4582.

\bibitem{[Eng]}M. Engelhardt, H. Reinhardt, Nucl. Phys. {\bf{B567}}
(2000) 249

\bibitem{[Reinh2]} H. Reinhardt, M. Engelhardt, Proceedings of the
XVIII Lisbon Autumn School, ``Topology of Strongly Correlated Systems'', Lisbon,
8-13 October, 2000, hep-th/0010031 

\bibitem{[Corn]}J. M. Cornwall, Phys. Rev. {\bf{D61}} (2000) 085012

\bibitem{[Eng3]} M. Engelhardt and H. Reinhardt, Nucl. Phys. {\bf{B585}}
(2000) 591

\bibitem{[Eng2]}M. Engelhardt, Nucl. Phys. {\bf{B585}} (2000) 614

\bibitem{[Ber2]}R. Bertle, M. Engelhardt, M. Faber, Phys. Rev. {\bf{D64}} (2001)
074504

\bibitem{[Fab]}M. Faber, J. Greensite, $\breve S$. Olejnik, Phys. Rev.
{\bf{D57}} (1998) 2603

\bibitem{[Lange]}K. Langfeld, H. Reinhardt, M. Quandt, hep-th/9610213

\bibitem{[Reinh]}H. Reinhardt, Nucl. Phys. {\bf{B503}} (1997) 505

\bibitem{[Ber]}R. Bertle, M. Faber, J. Greensite. $\breve S$. Olejnik, JHEP
{\bf{9903}} (1999) 019 

\bibitem{[Debbio]}L. Del Debbio. M. Faber, J. Greensite and $\breve S$. Olejnik,
in: {\it{New Developments in Quantum Field Theory}}, eds. P. H. Damgaard and J.
Jurkiewicz (Plenum Press, New York - London, 1998); hep-lat/9708023

\end{thebibliography}
\end{document}